\pdfoutput=1
\documentclass[usenatbib]{mnras}
\usepackage{newtxtext,newtxmath}
\usepackage{amsmath}
\usepackage{ctable}
\usepackage{verbatim}
\usepackage{url}
\usepackage{subcaption}
\usepackage{bm}
\usepackage[export]{adjustbox}

\newcommand{\be}{\begin{equation}}
\newcommand{\ee}{\end{equation}}

\newcommand{\etal}{et al.}

\newcommand{\msun}{\mathrm{M_{\sun}}}

\newcommand{\Alf}{{Alfv\'en}}

\newcommand{\ICsurl}{\href{http://www.tapir.caltech.edu/~phopkins/publicICs}{\url{http://www.tapir.caltech.edu/~phopkins/publicICs}}}
\newcommand{\FIREurl}{\href{http://fire.northwestern.edu}{\url{http://fire.northwestern.edu}}}
\newcommand{\gizmourl}{\href{http://www.tapir.caltech.edu/~phopkins/Site/GIZMO.html}{\url{http://www.tapir.caltech.edu/~phopkins/Site/GIZMO.html}}}

\newcommand{\Msun}{{M_{\odot}}\ }

\newcommand{\acknowledgments}{\begin{small}\section*{Acknowledgments}\end{small}}
\newcommand\altaffilmark[1]{$^{#1}$}
\newcommand\altaffiltext[1]{$^{#1}$}
\newcommand{\dataavailability}[1]{\begin{small}\section*{Data Availability}\end{small}{\noindent #1}\vspace{5pt}}

\title[Where oh where did my virial shock go?]{Virial shocks are suppressed in cosmic ray-dominated galaxy halos}

\author[Ji \etal]{
\parbox[t]{\textwidth}{ 
Suoqing Ji,\altaffilmark{1}
Du\v{s}an Kere\v{s},\altaffilmark{2}
T.~K.\ Chan,\altaffilmark{2,3}
Jonathan Stern,\altaffilmark{4}
Cameron B.~Hummels,\altaffilmark{1}
Philip F.~Hopkins,\altaffilmark{1}
Eliot Quataert\altaffilmark{5,6} and
Claude-Andr{\'e} Faucher-Gigu{\`e}re\altaffilmark{4}
}
\vspace*{6pt}
\\
\altaffiltext{1}{TAPIR, Mailcode 350-17, California Institute of Technology, Pasadena, CA 91125, USA. E-mail:suoqing@caltech.edu} \\
\altaffiltext{2}{Department of Physics, Center for Astrophysics and Space
Science, University of California at San Diego, 9500 Gilman Drive, La Jolla, CA
92093} \\
\altaffilmark{3}{Institute for Computational Cosmology, Durham University, South Road, Durham DH1 3LE, UK} \\
\altaffiltext{4}{Department of Physics and Astronomy and CIERA, Northwestern University, 2145 Sheridan Road, Evanston, IL 60208, USA} \\
\altaffiltext{5}{Department of Astronomy and Theoretical Astrophysics Center, University of California Berkeley, Berkeley, CA 94720} \\
\altaffiltext{6}{Department of Astrophysical Sciences, 4 Ivy Lane, Princeton
University, Princeton, NJ 08544} \vspace*{6pt}}

\date{Working Document}
\begin{document}
\maketitle
\label{firstpage}

\begin{abstract}
We study the impact of cosmic rays (CRs) on the structure of virial shocks,
using a large suite of high-resolution cosmological FIRE-2 simulations
accounting for CR injection by supernovae. In massive ($M_{\rm halo} \gtrsim
10^{11}\,M_{\sun}$), low-redshift ($z\lesssim 1-2$) halos, which are expected to
form ``hot halos'' with slowly-cooling gas in quasi-hydrostatic equilibrium
(with a stable virial shock), our simulations without CRs do exhibit clear
virial shocks. The cooler phase condensing out from inflows becomes
pressure-confined to over-dense clumps, embedded in low-density, volume-filling
hot gas whose cooling time is much longer than inflow time. The gas thus
transitions sharply from cool free-falling inflow, to hot and thermal-pressure
supported at approximately the virial radius ($\approx R_{\rm vir}$), and the
shock is quasi-spherical. With CRs, we previously argued that halos in this
particular mass and redshift range build up CR-pressure-dominated gaseous halos.
Here, we show that when CR pressure dominates over thermal pressure, there is no
significant virial shock. Instead, inflowing gas is gradually decelerated by the
CR pressure gradient and the gas is relatively subsonic out to and even beyond
$R_\mathrm{vir}$. Rapid cooling also maintains sub-virial temperatures in the
inflowing gas within $\sim R_\mathrm{vir}$.
\end{abstract}


\section{Introduction}
\label{sec:intro}

The thermal evolution of accreted gas onto galaxies is one of the essential
processes shaping galaxy formation. In the classic scenario, initially cold gas
from the intergalactic medium (IGM) is shock-heated when falling into dark
matter (DM) halos, and reaches quasi-hydrodynamical equilibrium at a virial
temperature $T_\mathrm{vir} \sim 10^6 (V_c/ 167\,\mathrm{km/s})^2$, where $V_c$
is the circular velocity. After that, hot gaseous halos start to cool from
inside out; the inner region contracts quasi-statically and feeds onto galactic
disks as the fuel for star formation
\citep{rees1977cooling,white1978core,fall1980formation}.
\citet{birnboim2003virial} explore this scenario into further details by
carrying out an analytic study with 1D spherical hydrodynamic simulations, where
the stability of radiative shocks in gas falling into DM halos is investigated.
It is shown that when cooling is inefficient, the incoming gas accelerated by
the DM gravitational potential accretes supersonically and evolves into shocks
as the gas piles up. The post-shock regions become thermalized and provide
thermal pressure support for newly accreted gas, and therefore a stable virial
shock front can form at virial radius $R_\mathrm{vir}$. In contrast, when
cooling is efficient, the post-shock regions rapidly radiates away its thermal
energy and cannot support a stable shock front. A general condition for stable
virial shocks is when the local gas cooling timescale is greater than the gas
compression time, which corresponds to a critical halo mass greater than $\sim
10^{11.5}\,\mathrm{M}_\odot$ with a metallicity of $\sim 0.05 Z_\odot$
\citep{dekel2006galaxy}. As a result, in massive halos of $M_\mathrm{halo}
\gtrsim 10^{11.5}\,\mathrm{M}_\odot$,  gas is shock-heated after entering galaxy
halos and ultimately accreted on to galaxies.

However, various complications emerge when spherical-symmetry is broken or
heating by feedback is included. A number of multi-dimensional numerical
simulations have extended this picture of gas accretion to account for
three-dimensional effects important in a realistic cosmological environment:
cold gas can possibly accrete in the form of anisotropic filamentary inflows
which penetrate hot halo gas and feed galaxies directly
\citep{kerevs2005galaxies}. Therefore, there might exist two distinct regimes
for gas accretion in DM halos, the ``cold'' mode in low-mass halos and the
``hot'' mode in high-mass ones at low redshifts respectively. At high redshift,
one can have efficient cold mode accretion even in massive halos that are
dominated by hot virialized atmospheres (e.g.
\citealt{kerevs2005galaxies,kerevs2009galaxies}). A number of numerical studies
utilizing cosmological simulations focus on the detailed transition between
these two regimes (e.g.,
\citealt{ocvirk2008bimodal,kerevs2009seeding,brooks2009role,oppenheimer2010feedback,faucher2011baryonic,van2011rates,nelson2013moving}),
and how galactic feedback from stars or AGN and hydrodynamic instabilities
affect this transition (e.g.,
\citealt{kerevs2009seeding,faucher2015neutral,faucher2016stellar,fielding2017impact,nelson2019first,stern2020virialization,esmerian2020thermal}).

In the meantime, a significant observational effort has been devoted to
disentangling the cold- and hot-mode accretion scenarios. For very large halo
masses corresponding to galaxy clusters ($\gtrsim 10^{14}\,\mathrm{M}_\odot$),
the virial temperature reaches $\gtrsim 10^7\,\mathrm{K}$ which corresponds to a
cooling time which is longer than the Hubble time. Hence hot gaseous halos are
expected, and these have been observed in X-ray emission
\citep{li2008chandra,anderson2011detection,fang2012hot} and Sunyaev-Zel'dovich
(SZ) effect \citep{planck13,anderson15}. On the other hand, evidence of
cold-mode accretion, expected to predominate in much lower-mass halos, is
relatively sparse due in great part to both low emissivities of cool halo gas
\citep{faucher2010lyalpha,van2013soft,sravan2016strongly}, and low covering
fractions of Lyman limit systems which trace cold accretion streams and thus are
hard to probe via absorption
\citep{faucher2011small,fumagalli2011absorption,faucher2015neutral,hafen2017low}.
Observational evidences have been significantly enriched by recent observations
of circumgalactic medium (CGM) \citep{tumlinson2017circumgalactic}. At high
redshifts, detected Ly$\alpha$ emission
\citep{cantalupo14,martin2015lyalpha,hennawi15,cai17} might suggest the
existence of cold filamentary inflows \citep{dijkstra2009lyalpha} or thermally
radiating cool gas \citep{fardal2001cooling}. At low redshifts, observations via
quasar absorption lines
\citep{stocke2013characterizing,werk2014cos,stern2016universal,prochaska2017cos}
agree on a total mass of $\gtrsim 10^{10}\,\mathrm{M}_\odot$ of cool gas at $\sim
10^4\,\mathrm{K}$ in MW-mass halos, though whether these observations indicate
the existence of a virial shock is unclear \citep{stern2018does}. Therefore, it
becomes particularly interesting to ask whether virial shocks can form in those
systems, and how the thermal state of halo gas is affected by the presence or
absence of virial shocks.

In recent years, the impact of non-thermal processes on the halo gas, such as
magnetic fields (e.g., \citealt{ji2018impact,van2020effect}) and cosmic rays
(CRs) (e.g.,
\citealt{salem2016role,farber2018impact,ji2020properties,butsky2020impact,buck2020effects}),
is under increasingly active investigation. \citet{ji2020properties} utilized
FIRE-2 simulations\footnote{\FIREurl} with cosmic ray (CR) physics incorporated
\citep{hopkins2020but,chan2019cosmic} to investigate the effects of CRs on the
CGM properties. \citet{hopkins2020but} and \citet{ji2020properties} demonstrated
that for certain assumptions about CR propagation, the CGM in MW-mass galaxy
halos can be supported by CR pressure rather than thermal pressure, which leads
to dramatic changes of the physical states of halo gas. \citet{ji2020properties}
found that in non-CR halos of MW mass ($\sim 10^{12} \,\mathrm{M}_\odot$), halo
gas is primarily warm-hot ($\gtrsim 10^{5}\,\mathrm{K}$) with highly
anisotropic, thermal pressure-confined cool filaments embedded, while in CR
pressure-dominated ones, halo gas is much cooler at a few $10^4\,\mathrm{K}$,
and the CR pressure-supported cool phase is diffuse and volume-filling over the
entire halos. The gas kinematics in CR-dominated halos are further discussed in
\citet{hopkins2020cosmic} with an emphasis on outflows. This work will primarily
focus on the impact of CRs on the properties of inflows, as well as its
consequence for the formation of virial shocks and the thermal state of halo
gas.

The outline of the paper is as follows: in \S\ref{sec:methods}, we describe our
computational methods. In \S\ref{sec:theory}, we describe theoretical
expectations, and in \S\ref{sec:results}, we present our findings. We conclude
in \S\ref{sec:conclusions}.

\begin{footnotesize}
\ctable[caption={{\normalsize Zoom-in simulation volumes run to $z=0$ (see
  \citealt{hopkins:fire2.methods} for details). All units are
  physical.}\label{tbl:sims}},center,star]{lcccccr}{\tnote[ ]{Halo/stellar
  properties listed refer only to the original ``target'' halo around which the
  high-resolution volume is centered: these volumes can reach up to $\sim
  (1-10\,{\rm Mpc})^{3}$ comoving, so there are actually several hundred
  resolved galaxies in total. {\bf (1)} Simulation Name: Designation used
  throughout this paper. {\bf (2)} $M_{\rm halo}^{\rm vir}$: Virial mass
  \citep[following][]{bryan.norman:1998.mvir.definition} of the ``target'' halo
  at $z=0$. {\bf (3)} $M_{\ast}^{\rm MHD+}$: Stellar mass of the central galaxy
  at $z=0$, in our non-CR, but otherwise full-physics (``MHD+'') run. {\bf (4)}
  $M_{\ast}^{\rm CR+}$: Stellar mass of the central galaxy at $z=0$, in our
  ``default'' (observationally-favored) CR+ ($\kappa=3e29$) run. {\bf (5)}
  $m_{i,\,1000}$: Mass resolution: the baryonic (gas or star) particle/element
  mass, in units of $1000\,\msun$. The DM particle mass is always larger by the
  universal ratio, a factor $\approx 5$. {\bf (6)} $\langle \epsilon_{\rm gas}
  \rangle^{\rm sf}$: Spatial resolution: the gravitational force softening
  (Plummer-equivalent) at the mean density of star formation (gas softenings are
  adaptive and match the hydrodynamic resolution, so this varies), in the MHD+
  run. Typical time resolution reaches $\sim 100-100\,$yr, density resolution
  $\sim 10^{3}-10^{4}\,{\rm cm^{-3}}$. {\bf (7)} Additional
  notes.}}{\hline\hline
Simulation & $M_{\rm halo}^{\rm vir}$ & $M_{\ast}^{\rm MHD+}$ & $M_{\ast}^{\rm CR+}$ & $m_{i,\,1000}$ & $\langle \epsilon_{\rm gas} \rangle^{\rm sf}$ & Notes \\
Name \, & $[\msun]$ &  $[\msun]$  &   $[\msun]$  &$[1000\,\msun]$ & $[{\rm pc}]$ & \, \\ 
\hline
{\bf m10q} & 8.0e9 & 2e6 & 2e6 & 0.25 & 0.8 & isolated dwarf in an early-forming halo \\
{\bf m11b} & 4.3e10 & 8e7 & 8e7 & 2.1 & 1.6 & disky (rapidly-rotating) dwarf \\
{\bf m11e} & 1.4e11 & 1e9 & 7e8 & 7.0 & 2.0 & low surface-brightness dwarf \\
{\bf m11d} & 3.3e11 & 4e9 & 2e9 & 7.0 & 2.1 & late-forming, ``fluffy'' halo and galaxy \\
{\bf m11f} & 5.2e11 & 3e10 & 1e10 & 12 & 2.6 & early-forming, intermediate-mass halo \\
{\bf m12i} & 1.2e12 & 7e10 & 3e10 & 7.0 & 2.0 & ``Latte'' halo, later-forming MW-mass halo, massive disk\\  
{\bf m12m} & 1.5e12 & 1e11 & 3e10 & 7.0 & 2.3 & earlier-forming halo, features strong bar at late times \\ 
{\bf m12f} & 1.6e12 & 8e10 & 4e10 & 7.0 & 1.9 & MW-like disk, merges with LMC-like companion \\ 
\hline\hline
}
\end{footnotesize}

\section{Methods}
\label{sec:methods}

The specific simulations studied here are the same as those presented and
studied in \citet{hopkins2020but} and \citet{ji2020properties}, where the
details of the numerical methods are described. We therefore only briefly
summarize here. The simulations were run with {\small GIZMO}\footnote{A public
version of {\small GIZMO} is available at \gizmourl} \citep{hopkins:gizmo}, in
its meshless finite-mass MFM mode (a mesh-free finite-volume Lagrangian Godunov
method). The simulations solve the equations of ideal magneto-hydrodynamics
(MHD) as described and tested in \citep{hopkins:mhd.gizmo,hopkins:cg.mhd.gizmo},
with anisotropic Spitzer-Braginskii conduction and viscosity as described in
\citet{hopkins:gizmo.diffusion,su:2016.weak.mhd.cond.visc.turbdiff.fx} and
\citet{hopkins2020but}. Gravity is solved with adaptive Lagrangian force
softening for gas (so hydrodynamic and force resolutions are matched). 

All our simulations include magnetic fields, anisotropic Spitzer-Braginskii
conduction and viscosity, and the physics of cooling, star formation, and
stellar feedback from the FIRE-2 version of the Feedback in Realistic
Environments (FIRE) project \citep{hopkins2014galaxies}, described in detail in
\citet{hopkins:fire2.methods}. Gas cooling is followed from $T=10-10^{10}\,$K
(including a variety of process, e.g.\ metal-line, molecular, fine-structure,
photo-electric, photo-ionization, and more, accounting for self-shielding and
both local radiation sources and the meta-galactic background with the FG09 UVB
model from \citet{faucher2009new} (see \citealt{hopkins:fire2.methods} for more
details). We follow 11 distinct abundances accounting for turbulent diffusion of
metals and passive scalars as in
\citet{colbrook:passive.scalar.scalings,escala:turbulent.metal.diffusion.fire}.
Gas is converted to stars using a sink-particle prescription if and only if it
is locally self-gravitating at the resolution scale \citep{hopkins:virial.sf},
self-shielded/molecular \citep{krumholz:2011.molecular.prescription},
Jeans-unstable, and denser than $>1000\,{\rm cm^{-3}}$. Each star particle is
then evolved as a single stellar population with IMF-averaged feedback
properties calculated following {\small STARBURST}99 \citep{starburst99} for a
\citet{kroupa:2001.imf.var} IMF and its age and abundances. We explicitly treat
mechanical feedback from SNe (Ia \&\ II) and stellar mass loss (from O/B and AGB
stars) as discussed in \citet{hopkins:sne.methods}, and radiative feedback
including photo-electric and photo-ionization heating and UV/optical/IR
radiation pressure with a five-band radiation-hydrodynamics scheme as discussed
in \citet{hopkins:radiation.methods}. Conduction adds the parallel heat flux
$\kappa_{\rm cond}\,\hat{\bm{B}}\,(\hat{\bm{B}}\cdot \nabla T)$, and viscosity
the anisotropic stress tensor $\Pi \equiv -3\,\eta_{\rm visc}\,(\hat{\bf
B}\otimes\hat{\bm{B}} - \mathbb{I}/3)\,(\hat{\bm{B}}\otimes\hat{\bm{B}} -
\mathbb{I}/3) : (\nabla\otimes{\bm{v}})$ to the gas momentum and energy
equations, where the parallel transport coefficients $\kappa_{\rm cond}$ and
$\eta_{\rm visc}$ follow the usual
\citet{spitzer:conductivity,braginskii:viscosity} form, accounting for
saturation following \citet{cowie:1977.evaporation}, and accounting for plasma
instabilities (e.g.\ Whistler, mirror, and firehose) limiting the heat flux and
anisotropic stress at high plasma-$\beta$ following
\citet{komarov:whistler.instability.limiting.transport,squire:2017.max.braginskii.scalings,squire:2017.kinetic.mhd.alfven,squire:2017.max.anisotropy.kinetic.mhd}.
The simulations are cosmological ``zoom-in'' runs with a high-resolution region
(of size ranging from $\sim 1$ to a few Mpc on a side) surrounding a ``primary''
halo of interest \citep{onorbe:2013.zoom.methods};\footnote{For the MUSIC
\citep{hahn:2011.music.code.paper} files necessary to generate all ICs here,
see:\\ \ICsurl} the properties of these primary halos (our main focus here, as
these are the best-resolved in each box) are given in Table~\ref{tbl:sims}.
Details of all of these numerical methods are in \citet{hopkins:fire2.methods}.

Our ``CRs'' or ``CR+'' simulations include all of the above, and add our ``full
physics'' treatment of CRs as described in detail in \citet{chan2019cosmic} and
\citet{hopkins2020but}. We evolve a ``single bin'' ($\sim$\,GeV) or constant
spectral distribution of CRs as an ultra-relativistic ($\gamma=4/3$) fluid,
accounting for injection in SNe shocks (with a fixed fraction $\epsilon_{\rm
cr}=0.1$ of the initial SNe ejecta kinetic energy in each time-resolved
explosion injected into CRs), collisional (hadronic and Coulomb) losses from the
CRs (with a fraction of this loss thermalizing and heating gas) following
\citet{guo.oh:cosmic.rays}, advection and adiabatic work (in the local ``strong
coupling'' approximation, so the CR pressure contributes to the total pressure
in the Riemann problem for the gas equations-of-motion), and CR transport
including anisotropic diffusion and streaming
\citep{mckenzie.voelk:1982.cr.equations}. We solve the transport equations using
a two-moment approximation to the full collisionless Boltzmann equation (similar
to the scheme in \citealt{jiang.oh:2018.cr.transport.m1.scheme}), with a
constant parallel diffusivity $\kappa_{\|}$ (perpendicular $\kappa_{\bot}=0$);
streaming velocity ${\bm{v}}_{\rm stream} = -v_{\rm stream}\,\hat{\bf
B}\,(\hat{\,\mathrm{M}_\odot}\cdot \hat{\nabla} P_{\rm cr})$ with $v_{\rm
stream}=3v_{A}$, the local \Alf\ speed
\citep{skilling:1971.cr.diffusion,holman:1979.cr.streaming.speed,kulsrud:plasma.astro.book,yan.lazarian.2008:cr.propagation.with.streaming};
and the ``streaming loss'' term ${\bm{v}}_{\rm A}\cdot \nabla P_{\rm cr}$
thermalized (representing losses to plasma instabilities at the gyro scale;
\citealt{wentzel:1968.mhd.wave.cr.coupling,kulsrud.1969:streaming.instability}).

Our ``baseline'' or ``no CRs'' simulations include all the physics above except
CRs: these are the ``MHD+'' simulations in \citet{hopkins2020but}. Note there we
also compared a set without magnetic fields, conduction, or viscosity (the
``Hydro+'' runs); but as shown
\citet{su:2016.weak.mhd.cond.visc.turbdiff.fx,hopkins2020but}, the differences
between ``Hydro+'' and ``MHD+'' runs are largely negligible, and we confirm this
here. Our default ``CR'' simulations adopt $\epsilon_{\rm cr}=0.1$, $v_{\rm
stream}=3v_{A}$, and $\kappa_{\|}=3\times10^{29}\,{\rm cm^{2}\,s^{-1}}$: these
are the ``CR+($\kappa=3e29$)'' simulations in \citet{hopkins2020but}. Although
we considered variations to all of these CR physics and, in particular, the
diffusivity (which is not known {\em a priori}) in \citet{hopkins2020but}, we
showed that the observational constraints from e.g.\ spallation and more
detailed measurements in the MW and $\gamma$-ray emission in local galaxies were
all consistent with the default ($\kappa_{\|}=3\times10^{29}\,{\rm
cm^{2}\,s^{-1}}$) model here.

\section{Theoretical Expectations}
\label{sec:theory}

Assume that we have a spherically-symmetric, steady-state inflow from
$r\rightarrow\infty$ into the center of a halo at $r\rightarrow 0$, in a halo
which is approximately quasi-hydrostatic over a timescale of $\sim
\mathrm{Gyr}$ at low redshifts. The continuity equation in steady state
($\partial \rho/\partial t = 0$) requires $4\pi\,\rho\,r^{2}\,v_{r} =
\dot{M}_{\rm in}=$\,constant as a function of radius, so we can replace $v_{r}
\rightarrow \dot{M}_{\rm in}/4\pi\,\rho\,r^{2}$. The momentum equation,
$r^{-2}\,\partial(\rho\,r^{2}\,v_{r}^{2})/\partial r = -\partial P/\partial r +
\rho\,\partial \Phi/\partial r$ can then be written:
\begin{align}
\label{eqn:infall.eqm}
\left(\frac{\dot{M}_{\rm in}}{4\pi\,\rho\,r^{2}} \right)^{2}\,\left( 2 + \frac{d\ln{\rho}}{d\ln r} \right) + \frac{1}{\rho}\,\frac{d P}{d\ln r} + V_{c}^{2}(r) = 0
\end{align}

\subsection{Without CRs}

The case without CRs where the gas is in hydrostatic equilibrium in a halo is
extensively studied in the literature, and we will only briefly mention it here
for reference.

Without CRs, assume the pressure is primarily gas thermal pressure. Firstly
consider a simple case where the gas should follow a single adiabat, i.e.\ $P
\propto \rho^{\gamma}$. At large radii when gas is falling in from the cosmic
web (before a shock), the gas is approximately isothermal at the temperature set
by photo-ionization equilibrium, so $P\approx \rho\,c_{s}^{2}$, which allows us
to write Eq.~\ref{eqn:infall.eqm} as $(\partial \ln \tilde{\rho}/\partial \ln
r)\,(1+\tilde{\rho}^{-2}) = 2 - (V_{c}/c_{s})^{2}$ with $\tilde{\rho} \equiv
4\pi\,c_{s}\,r^{2}\,\rho / \dot{M}_{\rm in}$ and circular velocity $V_c(r) =
\sqrt{GM(r)/r}$. When $V_{c} \gtrsim \sqrt{2}\,c_{s}$ (all cases of interest
here), for any reasonable potential profile (e.g. \citealt{hernquist:profile},
\citealt{nfw:profile} profiles), it is easy to verify that this equation has no
physical, smooth solutions that continuously connect $r\rightarrow 0$ to
$r\rightarrow \infty$. Furthermore it is easy to extend this to any adiabat
$P\approx P_{0}\,(\rho/\rho_{0})^{\gamma}$ and show the same for $V_{c} \gg
c_{s}$. For a form of $V_{c}$ which generically transitions from flat or rising
at small $r$ to falling at large $r$, the solutions for $\rho$ and $P$ tend to
diverge around this critical radius (i.e.\ $\sim R_{\rm vir}$).

This simply indicates that the gas must undergo some sort of shock or change in
entropy as it flows in; as is well-known, this does not necessarily imply a
``standing'' virial shock. As a rule, although shocks occur as the gas flows in
whenever the maximum halo circular velocity $V_{\rm max} \gg 10\,{\rm
km\,s^{-1}}$ (the isothermal temperature of the IGM), if the post-shock cooling
time $t_{\rm cool}$ is much less than the dynamical time $t_{\rm dyn}$ at the
shock radius ($\sim R_{\rm vir}$, giving $t_{\rm dyn} \sim 0.1\,t_{\rm
Hubble}$), the gas essentially free-falls onto the galaxy in a ``cold flow'' and
no stable, standing shock forms. On the other hand in halos with $M_{\rm halo}
\gtrsim 10^{11}\,M_{\sun}$, $t_{\rm cool} \gtrsim t_{\rm dyn}$ and a virial
shock tends to form covering most of the solid angle around the halo, even if a
significant fraction of the gas actually accreted by the galaxy comes in via
filaments that can ``punch through'' this hot halo.

\subsection{In CR-dominated halos}
\label{sec:theory_cr}

The case of particular interest is where CR pressure {\em dominates} over
thermal pressure in the CGM. A simple analytic model for this regime is
discussed in more detail in \citet{hopkins2020but}, as well as
\citet{ji2020properties,hopkins2020cosmic}, so we only briefly review here. For
a constant effective (angle-averaged) diffusivity $\tilde{\kappa}$ which is
large enough ($\tilde{\kappa} \gtrsim 10^{29}\,{\rm cm^{2}\,s^{-1}}$) and galaxy
central gas densities which are low enough (e.g.\ like MW and lower-mass
``normal'' galaxies at $z\sim 0$, but much lower than observed starburst systems
such as M82 or Arp220), the CRs escape and form a quasi-equilibrium,
diffusion-dominated steady-state energy-density profile with $P_{\rm cr} \approx
\dot{E}_{\rm cr}/(12\pi\,\tilde{\kappa}\,r)$ at $r < r_{\rm stream}$ and $P_{\rm
cr} \approx \dot{E}_{\rm cr}/(12\pi\,v_{\rm stream}\,r^{2})$ at $r > r_{\rm
stream}$, with $r_{\rm stream} \equiv \tilde{\kappa}/v_{\rm stream} \sim
\tilde{\kappa}/v_{A}(r_{\rm stream})$. Here $\dot{E}_{\rm cr} = \epsilon_{\rm
cr}\,\dot{E}_{\rm SNe} = \epsilon_{\rm cr}\,u_{\rm SNe}\,\dot{M}_{\ast}$ is the
CR injection rate. If the above conditions are met, this can dominate over gas
thermal pressure and be the dominant source of pressure support for gas from
radii outside the disk to $>R_{\rm vir}$, in halos which are sufficiently
massive ($\gtrsim 10^{11}\,M_{\sun}$; where SFR and therefore $\dot{E}_{\rm cr}$
is sufficiently large), star-forming and low-redshift ($z\lesssim 1-2$; where
densities are low enough for CRs to escape and the gas densities are
sufficiently low for CR pressure to dominate). 

So let us assume $P \approx P_{\rm cr} \approx \dot{E}_{\rm cr} /
(12\pi\,\tilde{\kappa}\,r\,[1+r/r_{\rm stream}])$ dominates the pressure. As
discussed in the papers above, this immediately defines a critical equilibrium
density profile $\rho_\mathrm{eq}$ where $dP/dr = \rho_\mathrm{eq} V_c^2/r$,
i.e. where CR pressure balances gravity: denser gas ($\rho > \rho_\mathrm{eq}$)
will sink, while less dense gas ($\rho < \rho_\mathrm{eq}$) will be accelerated
outward. Because the injection rate $\dot{E}_{\rm cr}$ is time-averaged over the
CR diffusion time out to $r$ ($\gtrsim$\,Gyr at the radii of interest), it is
not sensitive to short-timescale fluctuations in SFR. Moreover, because the CRs
diffuse rapidly on small scales through any structure, this profile is
essentially independent of perturbations to the local inflow rate and gas
density structure. Inserting this in Eq.~\ref{eqn:infall.eqm}, it is easy to
integrate numerically and verify (for any reasonable $V_{c}(r)$ profile) not
only that there are trivial solutions with constant inflow $\dot{M}_{\rm in}$,
but that these solutions are {\em smooth}, continuous, infinitely
differentiable, positive-definite, and monotonic in $\rho$ and $v_{r}$. 

If we assume the potential follows a \citet{hernquist:profile} or
\citet{nfw:profile} profile, then at small $r$, $V_{c}^{2} \propto r/r_{s}$
(where $r_{s}$ is a halo scale-length). If we also take $r \ll r_{\rm stream}$,
then the solutions become $\rho \rightarrow \rho_{\rm eq} = \dot{E}_{\rm
cr}/(12\pi\,\tilde{\kappa}\,r\,V_{c}^{2}(r)) \propto r^{-2}$, with $v_{r}
\rightarrow (3\,\dot{M}_{\rm in}\,\tilde{\kappa}/\dot{E}_{\rm
cr})\,(V_{c}^{2}/r) \approx\,$constant (so the inflow rate can be written:
$\dot{M}_{\rm in} = (\dot{E}_{\rm cr}\,v_{r} /
3\,\tilde{\kappa})\,(r/V_{c}^{2})$. Inserting $\epsilon_{\rm cr}=0.1$, the
value of $u_{\rm SNe}$ for our adopted IMF, $\tilde{\kappa}_{29} \equiv
\tilde{\kappa}/10^{29}\,{\rm cm^{2}\,s^{-1}}$, and the standard virial scalings
for dark matter halos for $r_{s}$ and $V_{c}(<r_{s})$, this gives $\dot{M}_{\rm
in} / \dot{M}_{\ast} \sim 0.6\,(1+z)^{-2}\,(v_{r}/10\,{\rm
km\,s^{-1}})\,(\Delta\Omega/4\pi)\,\tilde{\kappa}_{29}^{-1}\,(M_{\rm
halo}/10^{12}\,M_{\sun})^{-1/3}$ where $\Delta \Omega$ is the solid angle
covered by the inflow (assuming constant-$\Delta\Omega$). At large radii ($r \gg
r_{s},\,r_{\rm stream}$), if we assume $V_{c}^{2} \propto r^{-1}$ (e.g.\ a
Keplerian or \citealt{hernquist:profile} profile), then $\rho$ and $v_{r}$ both
gradually decline $\propto 1/r$ (our derivation does not include the Hubble flow
at large $r$, so this is the ``correct'' behavior).

In short, CR-dominated solutions {\em allow} for smooth quasi-isothermal inflow
of gas from infinity into the halo, {\em without a virial shock}. These
solutions have $\rho\propto v_{r} \propto r^{-1}$ at large radii and
$\rho\propto r^{-2}$, $v_{r} \sim v_{\rm terminal} \sim $\,constant at small
radii, without a discontinuity in the gas properties. They feature inflow rates
comparable to or larger than galaxy SFRs for relatively modest (transonic or
mildly super-sonic) inflow velocities.\footnote{It is also worth pointing out
that the infall rate and SFR are not independent of each other. Instead, the
infall rate is required to be least equal to SFR, or higher to account for
outflows.}

\section{Results}
\label{sec:results}

We begin by focusing on a representative case study of one MW-like,
$10^{12}\,\mathrm{M}_\odot$ halo ({\bf m12i}) from our suite. In \citet{hopkins2020but} and
\citet{ji2020properties}, we show that at $z=0$, the CR+ run of {\bf m12i}
studied here has CR pressure much larger than thermal pressure in CGM gas at
radii $\lesssim 500\,\mathrm{kpc}$, making this an excellent example of a ``CR
pressure-dominated halo''. In \S\ref{sect:halo_depd} below, we examine
additional halos to explore mass and environmental dependencies.

\subsection{Spatial structure}

\begin{figure*}
\begin{centering}
  \includegraphics[width={\textwidth}]{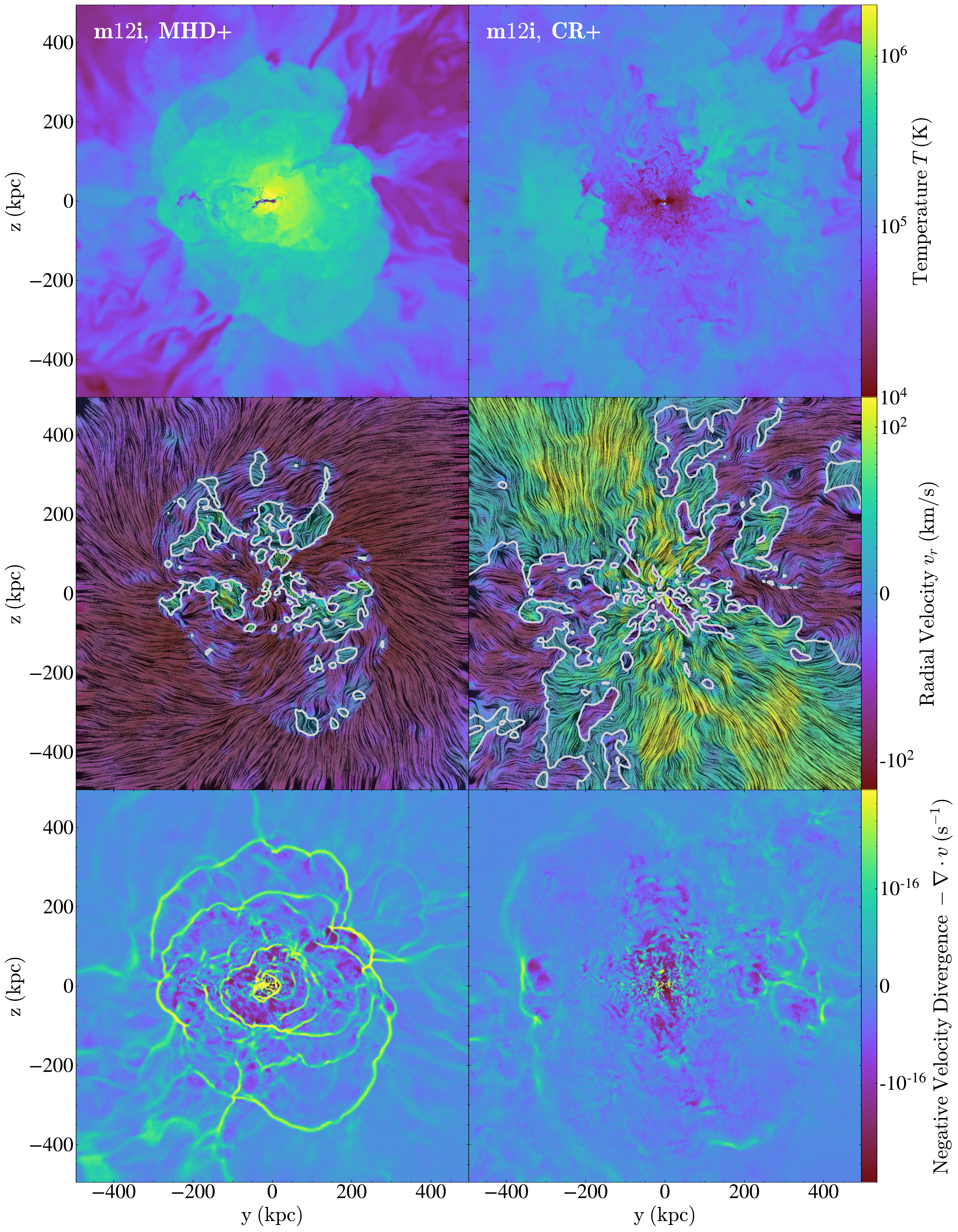}
\caption{Slice plots edge-on to the disk of (\emph{top}) gas temperatures,
(\emph{middle}) radial velocities superimposed with black velocity streamlines
and white contours at $v_r=0$), and (\emph{bottom}) negative velocity
divergences, in ``MHD+'' (\emph{left}) and ``CR+'' (\emph{right}) runs for {\bf
m12i} halo at $z=0$. MHD+ features a hot halo with clear shock fronts peaking
$-\nabla \cdot \bm{v}$, with the CGM dominated by inflows and trapped outflows. In the
CR+ run we see a much cooler halo, with extended outflows and much less apparent
shocks except for mild compression at $r\sim R_\mathrm{vir}$.
\label{fig:slices_m12i}}
\end{centering}
\end{figure*}

Fig. \ref{fig:slices_m12i} shows slice plots (edge-on to the disk) taken from
the snapshot at $z=0$ from the MHD+ and CR+ runs of {\bf m12i} halo. In the top
panel, MHD+ unambiguously produces a hot halo at $T\gtrsim 10^6\,\mathrm{K}$,
with virial shock fronts at $r\sim 300\,\mathrm{kpc}$ indicated by temperature
jumps. However, such a hot halo is absent from CR+; instead, a low-temperature
``void'' at $T\sim$ a few $10^4\,\mathrm{K}$ forms roughly aligned with bipolar
directions, surrounded by warm ambient medium of $\sim 10^5\,\mathrm{K}$ around
$r\sim 200\,\mathrm{kpc}$. 

The velocity structure is shown in the middle panel of Fig.
\ref{fig:slices_m12i}, where the velocity streamlines are plotted in black lines
and the background color represents the magnitude of the radial velocity $v_r$,
with white contours at values of $v_r=0$. MHD+ shows inflow in all directions
from the cosmic web onto a clear and large-scale coherent ($\gtrsim
100\,\mathrm{kpc}$) virial shock at $r\sim 300\,\mathrm{kpc}$, with a turbulent,
inflow-dominated halo interior to this. In contrast, CR+ shows inflow
penetrating the mid-plane down to $r\sim 100\,\mathrm{kpc}$ and strong bipolar
outflow extending beyond $500\,\mathrm{kpc}$, with no apparent virial shock. It
is worth noting that by comparing the velocity and temperature plots, we can see
that in CR+, the mid-plane inflow appears co-spatial with the warm
($10^5\,\mathrm{K}$) region, and the bipolar outflow corresponds to the cool (a
few $10^4\,\mathrm{K}$) gas. Detailed analysis on inflow/outflow properties will
be discussed later in \S\ref{sec:inoutflow}.

The shock structure can be observed clearly by plotting the negative of the
velocity divergence $-\nabla \cdot \bm{v}$, as shown in the bottom panel in Fig.
\ref{fig:slices_m12i}, where large positive values indicate converging flow at
shock fronts. For MHD+, inflowing streams are weakly compressed until they reach
$r\sim 300\,\mathrm{kpc}$ where the first shock front forms and encloses the
entire halo. Inside the shock-enclosed halo, a series of additional shocks are
triggered at $r < 300\,\mathrm{kpc}$ by the interaction between inflows and
outflows. In contrast, for CR+, no spatially coherent shock structure is
observed, and the gas is only mildly compressed within limited regions around
$r\sim 400\,\mathrm{kpc}$. Within $r \sim 100\,\mathrm{kpc}$ in CR+, relatively
large fluctuations in $\nabla \cdot \bm{v}$ appear due to supersonic turbulence,
as discussed later in \S\ref{sec:inflows}.

In short, a prominent, solid-angle covering virial shock is absent from
CR-dominated halos; the halo gas is cooler at a few $10^4\,\mathrm{K}$ with
extended bipolar outflows. In contrast, shocks are prevalent in non-CR halos
around and within the virial radius which heats up the halo gas to the virial
temperature; the halo is inflow-dominated, and outflows are sparse and trapped
inside the virial radius (the latter point is discussed in details in
\citealt{hopkins2020cosmic}). Compared to the MHD+ case, the CR+ run shows
highly anisotropic spatial structures \citep{qu2019warm}. 

\subsection{Properties of inflows and outflows}
\label{sec:inoutflow}

\begin{figure*}
  \begin{centering}
    \includegraphics[width={\textwidth}]{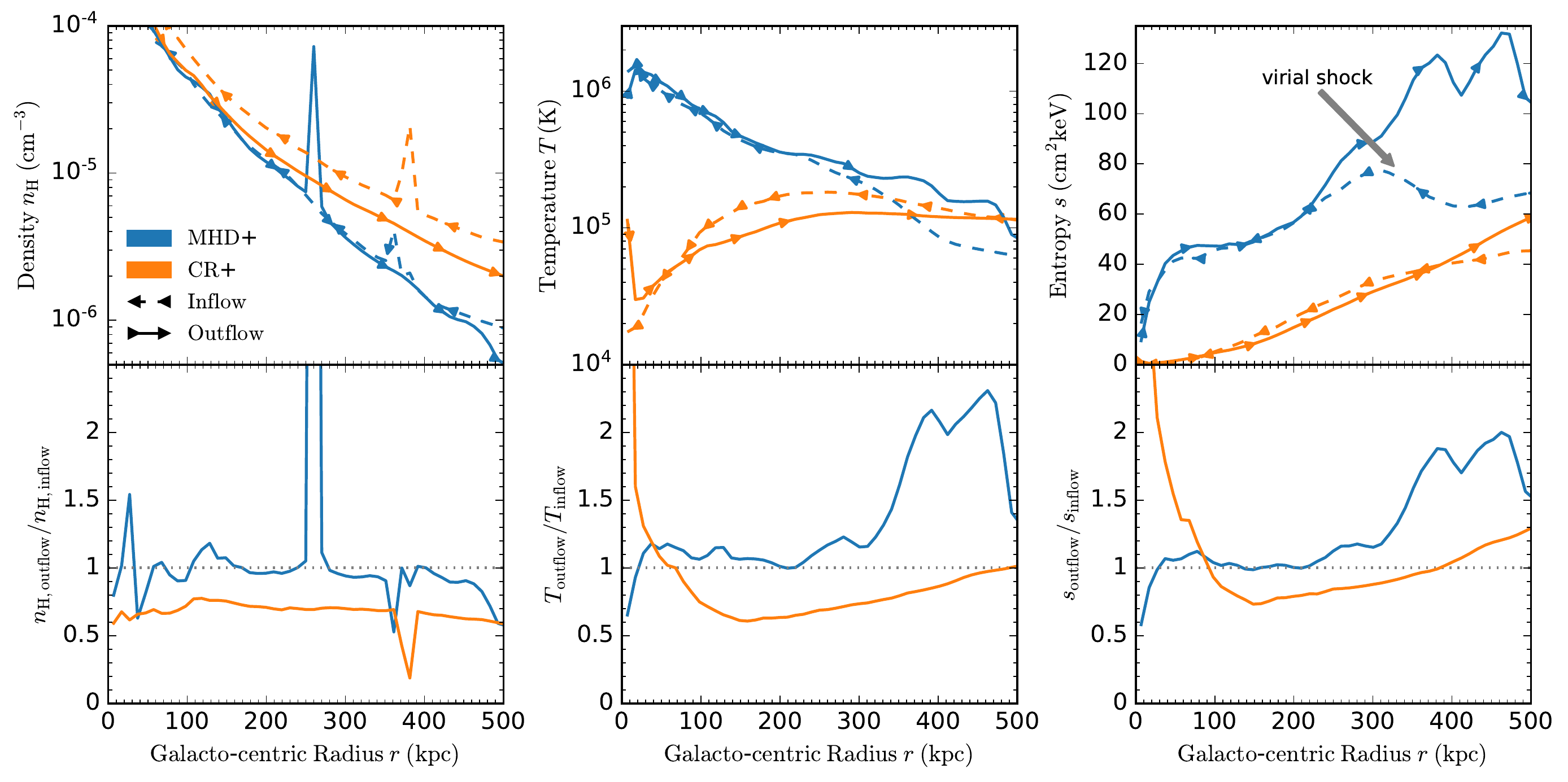}
  \end{centering}
    \caption{Top: volume-averaged radial profiles of gas number density
  (\emph{left}), temperature (\emph{middle}) and entropy (\emph{right}) of
  inflows (\emph{dashed with leftward arrows}) and outflows (\emph{solid with
  rightward arrows}) for the MHD+ (\emph{blue}) and CR+ (\emph{orange}) runs of
  the {\bf m12i} halo at $z=0$. Bottom panels show the ratios of the inflow and
  outflow values. Plotted values are volume-averaged in spherical shells at
  galacto-centric radius $r$. In MHD+, outflows have higher temperature and
  entropy than inflows beyond the virial shock ($r\gtrsim 300\,\mathrm{kpc}$),
  while inside the virial shock ($r\lesssim 300\,\mathrm{kpc}$) the density,
  temperature and entropy of inflows and outflows are similar. In CR+, the
  density, temperature and entropy of outflows are lower than inflows at most
  radii by a factor of a couple; the temperature and entropy of \underline{both}
  inflows and outflows are lower as well. Note the ``spikes'' in $n_\mathrm{H}$
  at certain $r$ owe to the exact location of a satellite galaxy in this
  snapshot. \label{fig:profile.flow.m12i}}
\end{figure*}

\begin{figure}
  \begin{centering}
    \includegraphics[width={\columnwidth}]{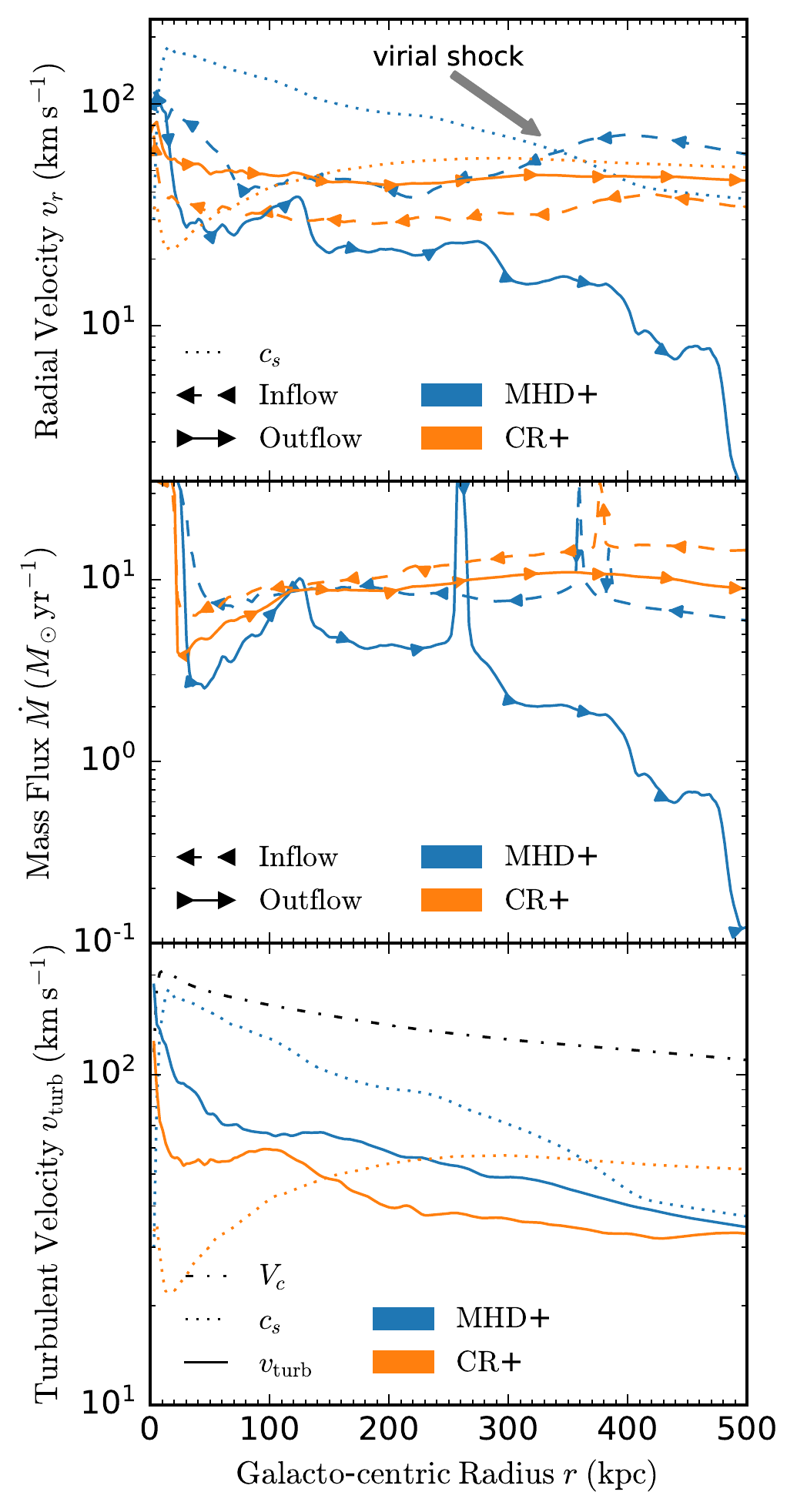}
  \end{centering}
    \caption{Radial profiles of radial velocities (\emph{top}), and mass fluxes
  (\emph{middle}) and turbulent velocities (\emph{bottom}) of inflows
  (\emph{dashed with leftward arrows}) and outflows (\emph{solid with rightward
  arrows}) for the MHD+ (\emph{blue}) and CR+ (\emph{orange}) runs of the {\bf
  m12i} halo at $z=0$ (as Fig. \ref{fig:profile.flow.m12i}). Thermal sound
  speeds $c_s$ (dotted) and circular velocities $V_c$ (dash-dotted) are also
  shown for reference. MHD+ is dominated by inflows while the outflows are hot
  (Fig. \ref{fig:profile.flow.m12i}) but slow and $\dot{M}$, $v_r$ drop rapid
  with $r$, indicating trapping. In CR+, both inflows and outflows have roughly
  constant fluxes and velocities with radius (i.e. are not trapped or
  decelerated); outflows are faster, but inflows carry slightly larger mass
  flux. The $\dot{M}_\mathrm{in}$ profile is quite consistent with the
  CR-dominant $\dot{M}$ solutions discussed in \S\ref{sec:theory_cr}; requiring
  $\dot{M}_\mathrm{in} \sim \dot{M}_\mathrm{out} + \dot{M}_\ast$ further gives
  $\dot{M}_\mathrm{out} \sim \frac{1}{2} \dot{M}_\mathrm{in}$, as shown in the
  middle panel. Note the ``spikes'' in $\dot{M}$ at certain $r$ owe to the exact
  location of a satellite galaxy in this snapshot. Turbulence is subsonic in
  MHD+ and at large radii ($r\gtrsim 100\,\mathrm{kpc}$) in CR+, but transits to
  supersonic in the inner halo ($r\lesssim 100\,\mathrm{kpc}$) of CR+ due to the
  decrease of the sound speed towards smaller radii, as shown in the bottom
  panel. Here we estimate turbulent velocities by computing velocity dispersions
  of $v_\theta$ and $v_\phi$ while excluding $v_r$ in spherical coordinates,
  since $\delta v_r$ is dominated by the shear motion between coherent inflows
  outflows which is actually not turbulence. \label{fig:profile.massflow.m12i}}
\end{figure}

To probe the shock structure quantitatively, Fig. \ref{fig:profile.flow.m12i}
presents the density, temperature and entropy profiles respectively for inflows
and outflows in the MHD+ and CR+ runs. Fig. \ref{fig:profile.massflow.m12i}
shows the inflow/outflow radial velocities, mass fluxes and turbulent
velocities.

\subsubsection{Outflows}

\citet{hopkins2020cosmic} present a detailed study of the effect of CRs on
outflows in these simulations, so we only briefly summarize outflows here (as
they are relevant for some inflow behaviors). In MHD+, outflows are
quasi-spherical; they have similar density compared to inflows but higher
temperature/entropy/thermal pressure (especially outside $\sim R_{\rm vir}$);
and the outflow temperature, radial velocity, and mass flux all decline with
increasing radius $r$. Outflow velocities are also sub-sonic and sub-virial, and
the outflows appear well-mixed with inflows well inside of $R_{\rm vir}$. All of
these are generic properties expected of trapped thermal-pressure-driven
outflows.

In CR+, outflows are bipolar and are not well-mixed with inflows: the outflow
velocity $v_{r}$ and mass flux $\dot{M}_{\rm out}$ are approximately constant
with radius (not trapped), with $\dot{M}_{\rm out} \sim
2\,\dot{M}_{\ast}\,(v_{r}/30\,{\rm km\,s^{-1}})$ in good agreement with the
analytic CR-pressure-dominated inflow solutions from \S~\ref{sec:theory} (given
the values $\tilde{\kappa}_{29}\sim1$, $M_{\rm halo} \sim 10^{12}\,M_{\odot}$,
$z\sim 0$, $\Delta \Omega/4\pi \sim 1/2$ appropriate for this snapshot).
Outflows are approximately isothermal and have lower temperature/entropy/thermal
pressure compared to inflows, indicating they are driven by non-thermal (CR)
pressure. Finally in CR+, outflows have lower densities, higher velocities, and
similar mass fluxes compared to inflows: all of these can again be predicted
from our simple analytic model (see \S~\ref{sec:theory}): gas with $\rho <
\rho_{\rm eq}$ is accelerated outward by CR pressure while gas with $\rho >
\rho_{\rm eq}$ sinks (since gravity exceeds CR pressure forces), so $\rho_{\rm
outflow} < \rho_{\rm eq} < \rho_{\rm inflow}$ should be true at all $r$ (and
indeed we find $\langle \rho_{\rm outflow} \rangle /\langle \rho_{\rm inflow}
\rangle \sim 0.6$ at all $r$). Moreover, as shown in \citet{hopkins2020cosmic},
a parcel of gas with some initial density $\rho_{i} \ne \rho_{\rm eq}$ (so CR
and gravity forces are imbalanced) is accelerated to a terminal speed $|v_{r}|
\propto (\rho_{i}/\rho_{\rm eq}[r_{i}])^{-1/2}$: so we should expect $v_{\rm
outflow}/v_{\rm inflow} \sim (\rho_{\rm outflow}/\rho_{\rm inflow})^{-1/2} \sim
1.3 > 1$. Using $\dot{M} \propto \rho\,v_{r}$, we then have $|\dot{M}_{\rm
outflow}|/|\dot{M}_{\rm inflow}| \propto (\rho_{\rm outflow}/\rho_{\rm
inflow})^{1/2} \sim 0.8 < 1$.

\subsubsection{Inflows}
\label{sec:inflows}

In MHD+, beginning from $r>500\,$kpc, inflows gradually accelerate to larger
$|v_{r}|$ as they fall to smaller $r$ (Fig.~\ref{fig:profile.massflow.m12i}),
maintaining the approximately isothermal IGM temperature (slightly decreasing
their entropy as $\rho$ increases; Fig.~\ref{fig:profile.flow.m12i}), until they
suddenly de-celerate and increase in temperature and entropy at $r \sim
300\,{\rm kpc} \sim R_{\rm vir}$ (the clear position of the virial shock in
Fig.~\ref{fig:slices_m12i}). Inside of $R_{\rm vir}$, $\dot{M}_{\rm in}$ remains
approximately constant, with inflows and outflows at similar $\rho$, $T$, $s$
(the result of efficient inflow/outflow mixing), with densities and temperatures
increasing and entropies decreasing at small $r$ and sub-sonic $v_{r}$,
indicative of a classic quasi-hydrostatic cooling-flow (i.e. ``hot halo'')
solution (see e.g.\ \citealt{stern2020virialization}), for more discussion of
these solutions in our non-CR runs).

The CR+ run, in contrast, does not exhibit these signatures. The inflow velocity
$v_{r}\sim 30-40\,{\rm km\,s^{-1}}$ and temperature $T\sim 10^{5}\,$K remain
approximately constant from $r \gtrsim 500\,$kpc to $r \ll 100\,$kpc, while
entropy $s$ gradually declines as $n$ gradually rises at smaller $r$. As
detailed above, the inflowing gas is systematically more dense and slower than
outflows (with slightly higher $\dot{M}$), as expected from analytic
steady-state models of gas in a CR-pressure-dominated halo. As the gas is
cooler, inflow velocities are trans-sonic at most radii. As discussed in
\citet{ji2020properties}, gas densities of both inflows and outflows are higher,
especially at large $r$, compared to the MHD+ run, as the additional CR pressure
can simply support more gas against gravity. Given the $z\sim 0$ SFR of this
galaxy $\dot{M}_{\ast} \sim 2-3\,M_{\odot}\,{\rm yr^{-1}}$ and $v_{r}$, the
inflow rate $\dot{M}_{\rm inflow}$ agrees remarkably well with our
order-of-magnitude scaling for a steady-state CR-pressure dominated inflow in
\S~\ref{sec:theory}.

In the left panel of Fig. \ref{fig:profile.flow.m12i}, the overall gas density
in CR+ is higher than that in MHD+ by a factor of a few, and notably, the
density outflow-to-inflow ratios in MHD+ and CR+ differs distinctly: in MHD+,
the inflow and outflow densities are similar, but in CR+, the density of
outflows is systematically lower than inflows with the ratio
$n_\mathrm{H,outflow} / n_\mathrm{H,inflow} \sim 0.6 $ -- $ 0.7 < 1$, which
holds for a wide spacial range over $500\,\mathrm{kpc}$. This is consistent with
previous findings in \citet{ji2020properties} and our theoretical expectations:
gas with $\rho < \rho_\mathrm{eq}$ is accelerated outward by CR pressure
gradients, while gas with $\rho > \rho_\mathrm{eq}$ falls towards the galactic
center since gravity exceeds CR pressure support locally, therefore the order
$\rho_\mathrm{outflow} < \rho_\mathrm{eq} < \rho_\mathrm{inflow}$ is always
valid.

The middle panel of Fig. \ref{fig:profile.flow.m12i} shows temperature profiles.
In MHD+, the temperatures of both inflows and outflows increase with decreasing
radius; outflows are warmer than inflows by a factor of $2$ at large radii of
$r\gtrsim300\,\mathrm{kpc}$. At smaller radii of $r\lesssim 300\,\mathrm{kpc}$,
the temperatures of inflows and outflows are similar. This is consistent with
the virial shock picture in non-CR halos: cooler inflows hit onto the virial
shock front at $r\sim 300\,\mathrm{kpc}$, get heated up by dumping shock kinetic
energy into thermal energy, and become fully-mixed into highly turbulent gas
inside the virial shock where the inflows and outflows have the same
temperature. Such an apparent signature of shocks is absent from the CR+ run:
both inflows and outflows are nearly isothermal at $T\sim 10^5\,\mathrm{K}$ for
$r\gtrsim 150\,\mathrm{kpc}$; outflows are cooler than inflows at most radii, in
contrast to MHD+ where outflows are hotter at large radii. The difference in
temperatures between inflows and outflows suggests the mixing of inflows and
outflows is inefficient in CR+, which can also be seen from Fig.
\ref{fig:slices_m12i} as well, where inflows and bi-conical outflows trace
different paths with clear boundaries between each other without significant
mixing.

In the right panel of Fig. \ref{fig:profile.flow.m12i}, the entropy of inflows
in MHD+ shows a bulge at $r\sim300\,\mathrm{kpc}$ due to virial shock heating,
and the entropy of outflows is $\sim 2$ times higher than inflows at large
radii. At $r\lesssim 200,\mathrm{kpc}$, inflows and outflows have the same
entropy, indicating efficient mixing within shocked halos. In CR+, the entropy
of both inflows and outflows monotonically and smoothly decrease towards smaller
radii, without shock-induced entropy input. Inside CR-dominated halos at
$100\,\mathrm{kpc} \lesssim r \lesssim 400\,\mathrm{kpc}$, outflows carry lower
entropy than inflows.

After examining the fluid properties of inflows and outflows respectively in
both MHD+ and CR+, we find that CRs dramatically change gas dynamics in galaxy
halos as well as the existence of virial shocks. In non-CR halos, cool gas
carrying low entropy flows inwards until it hits onto the virial shock front at
$r\sim300\,\mathrm{kpc}$, where it meets hot, high-entropy outflowing gas and is
heated by shock kinetic energy dissipation. Cool inflows and hot outflows are
fully mixed with each other and reach the same temperature and entropy inside
the virial shock front; there is no significant difference in gas densities
between outflows and inflows in non-CR halos. However, CR-dominated halos behave
in an opposite way by featuring cooler, lower-density and lower-entropy outflows
along bi-polar directions compared with inflows. Consequently, in CR-dominated
halos, inflows and outflows are not in thermal pressure equilibrium; instead,
outflows are under-pressured relative to inflows with
$P_\mathrm{thermal,outflow} < P_\mathrm{thermal,inflow}$, but still in total
pressure equilibrium with CR pressure taken into account, i.e.
$P_\mathrm{thermal} (r)+ P_\mathrm{cr} (r) \approx \text{constant}$ at a given
$r$ \citep{ji2020properties}. This suggests that CRs are more abundant in
bi-conical outflow regions than inflow regions.

As previously discussed, in CR-dominated halos, gas with density $\rho_i$
deviating from $\rho_\mathrm{eq}$ feels an imbalance between CR pressure
gradients and gravity, and thus is subject to local acceleration. Such a parcel
of gas with density $\rho_i$ is accelerated to a terminal speed of $v_r \propto
(\rho_i/\rho_\mathrm{eq})^{-1/2}$ \citep{hopkins2020cosmic}. Then given
$\rho_\mathrm{outflow}/\rho_\mathrm{inflow}\sim 0.6$ -- $0.7$ observed in Fig.
\ref{fig:profile.flow.m12i}, we have $|v_\mathrm{outflow}| /
|v_\mathrm{infow}|\propto(\rho_\mathrm{outflow}/\rho_\mathrm{inflow})^{-1/2} > 1$,
and mass flux $|\dot{M}_\mathrm{outflow}|/|\dot{M}_\mathrm{inflow}|\propto
(\rho_\mathrm{outflow}/\rho_\mathrm{inflow})^{1/2} < 1$ where $\dot{M}\propto
\rho v_r$. These ratios turn out to be consistent with profiles plotted in Fig.
\ref{fig:profile.massflow.m12i}, where for CR+ (orange lines), radial velocity
(top panel) is outflow-dominated, while mass flux (middle panel) is
inflow-dominated. And for MHD+ (blue lines in Fig.
\ref{fig:profile.massflow.m12i}), both radial velocity and mass flux are
dominated by inflows; radial velocity and mass flux associated with outflows
rapidly decline with radius, indicating outflows are significantly trapped. In
addition, due to the decrease of temperature at small radii of $r \lesssim
100\,\mathrm{kpc}$ in CR+ (middle panel of Fig. \ref{fig:profile.flow.m12i}),
the local sound speed drops below turbulent velocities and supersonic turbulence
develops, as shown in the bottom panel of Fig. \ref{fig:profile.massflow.m12i}.
The supersonic turbulence in the inner-halo region is reflected by relatively
large $\nabla \cdot v$ fluctuations at small radii of the CR-dominated halo
(bottom panel of Fig. \ref{fig:slices_m12i}).

We notice that due to the presence of CR pressure gradients, a \emph{much lower}
inflow velocity in CR+ than in MHD+ even outside $R_\mathrm{vir}$ is predicted
by our analytic model and demonstrated by our simulations, which leads to two
consequences. First, the inflow velocity (orange dashed line in the top panel of
Fig. \ref{fig:profile.massflow.m12i}) is already significantly \emph{subsonic}
at $r\gtrsim R_\mathrm{vir}$, therefore a virial shock is not expected to
develop; while in MHD+, inflows (blue dashed line in the top panel of Fig.
\ref{fig:profile.massflow.m12i}) travel at a speed exceeding the sound speed and
thus is \emph{supersonic} outside $R_\mathrm{vir}$, until they hit onto virial
shocks which raise local sound speed and become subsonic. Second, the inflow
velocity in CR+ is reduced as if the gas is falling into a much lower potential,
which effectively shifts the halo mass scale to the level characteristic of
$M_\mathrm{halo} \sim 10^{11}\,\mathrm{\Msun}$ where gas would be heated to
$\sim 2\times 10^5\,\mathrm{K}$, if shocked. Therefore even if gas did shock,
cooling at this temperature range would be too efficient to establish a stable
atmosphere. We discuss the efficiency of gas cooling in more detail below.

\subsection{Conditions for the formation of a stable virial shock}

We discuss a sufficient condition which accounts for the absence of virial
shocks in our  CR+ run. Since CR heating in the CGM is negligible
\citep{ji2020properties}, the leading terms governing gas thermal energy
$E_\mathrm{thermal}$ are adiabatic $PdV$ work applied to gas and radiative
cooling: $dE_\mathrm{thermal}/dt \sim P dV/dt - \int n^2 \Lambda dV$, where $dV$
is the volume element and $\Lambda$ is the cooling coefficient. A viral shock
cannot form if gas is net cooling, which in spherical symmetry we can write as
$dE_\mathrm{thermal} / dt \sim 4 \pi r^2 (P v_r - \int n^2 \Lambda dr) \lesssim
0$. Defining the cooling time and inflow time as $t_\mathrm{cool} \equiv 3/2
kT/n\Lambda$ and $t_\mathrm{inflow} \equiv r/v_r$, we can see this is equivalent
to $t_\mathrm{cool}\lesssim t_\mathrm{inflow}$. When this condition is
satisfied, inflowing gas can radiate away its thermal energy before accreting,
and thus remain cool without creating a shock. We note that the definition of
$t_\mathrm{cool}$ here does not include any heating terms, such as
photon-heating, CR heating, turbulent heating or CR adiabatic heating/cooling,
since $t_\mathrm{cool}$ is used to evaluate how quickly gas cools when the gas
deviates from its equilibrium temperature which is co-determined by gas cooling
and heating sources. We briefly discuss additional heating terms in more detail
in \S\ref{sect:heat_cool}.

\begin{figure*}
  \begin{centering}
    \includegraphics[width={0.9\textwidth}]{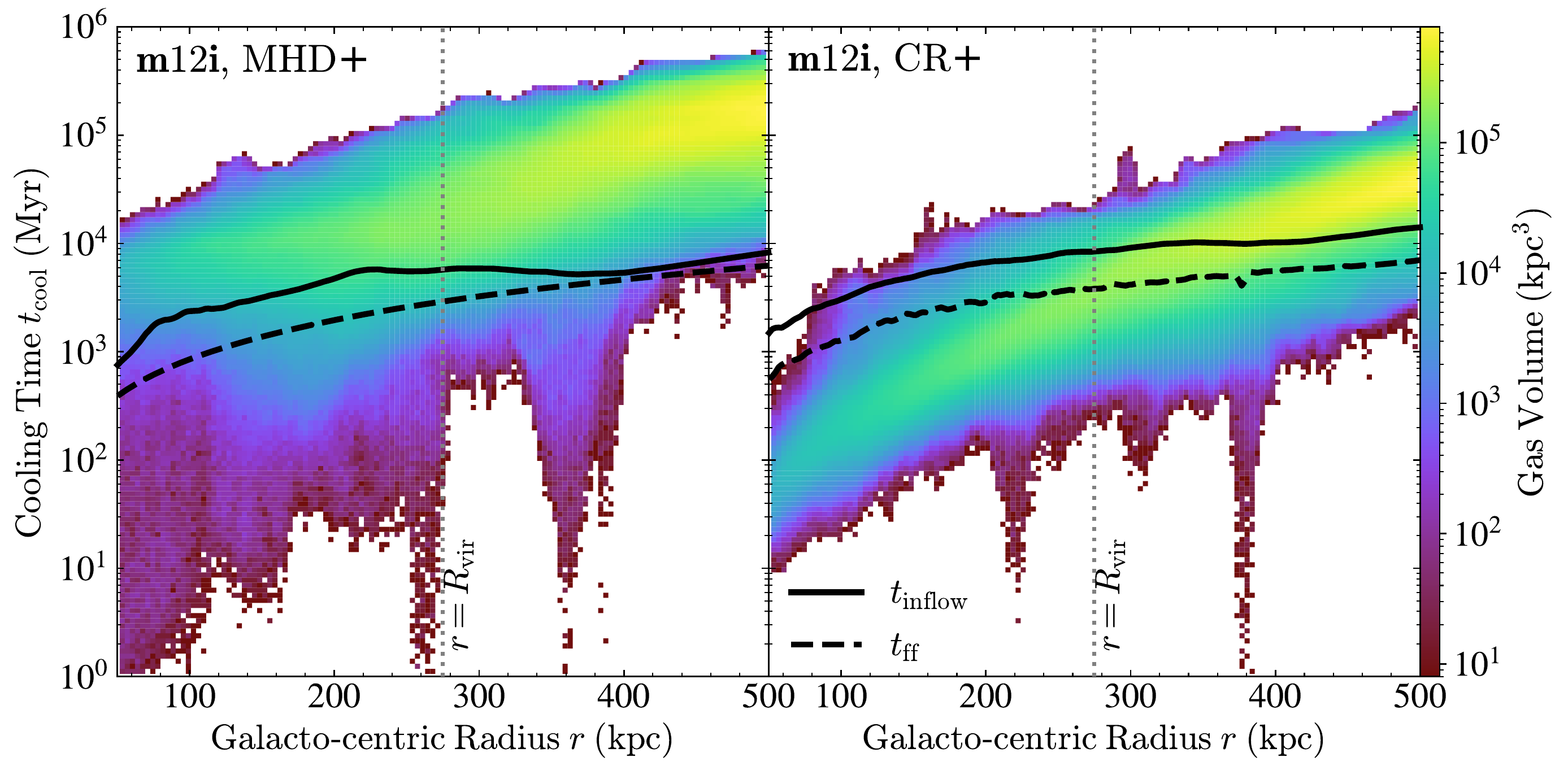}
  \end{centering}
  \caption{Radial profiles of cooling time $t_\mathrm{cool} =3/2\, k T/ n
  \Lambda$ (\emph{color}) weighted by gas volume, superposed with profiles of
  inflow time $t_\mathrm{inflow} = r / v_\mathrm{inflow}$ (\emph{solid}) and
  effective freefall time $t_\mathrm{ff} = \sqrt{2r/(\nabla \Phi+\nabla
  P_\mathrm{cr}/\rho)}$ (\emph{dashed}) in the MHD+ (\emph{blue}) and CR+
  (\emph{orange}) runs of the {\bf m12i} halo at $z=0$, where the location of
  $R_\mathrm{vir}$ marked with a dotted vertical line. The cooling time of the
  volume-filling phase in MHD+ is longer than CR+ by one order of magnitude or
  more. Compared to MHD+, CR pressure support in the CR+ run slightly increases
  $t_\mathrm{ff}$ and $t_\mathrm{inflow}$, but the large temperature difference
  (Fig. \ref{fig:profile.flow.m12i}) leads to much lower $T$ and hence
  $t_\mathrm{cool}$ in CR+, making $t_\mathrm{cool} \lesssim
  t_\mathrm{inflow},t_\mathrm{ff}$ for $r\lesssim 300$--$400\,\mathrm{kpc}$ in
  CR+ (while $t_\mathrm{cool} \gtrsim t_\mathrm{inflow}$ within $R_\mathrm{vir}$
  and $t_\mathrm{cool} \gg t_\mathrm{inflow},t_\mathrm{ff}$ outside
  $R_\mathrm{vir}$ in MHD+). \label{fig:profile.timescales.m12i}}
\end{figure*}

Fig. \ref{fig:profile.timescales.m12i} shows radial profiles of cooling time
$t_\mathrm{cool}$ (colored, weighted by gas volumes) and inflow time
$t_\mathrm{inflow}$ (solid) in our {\bf m12i} runs. Since CR pressure is
decoupled from gas thermal pressure and resists infall, we define an effective
freefall time $t_\mathrm{ff}\equiv \sqrt{2r/(\nabla \Phi+\nabla
P_\mathrm{cr}/\rho)}$ (with $\Phi$ the gravity potential) to include the effects
of CRs. For MHD+, the relation $t_\mathrm{cool} \gtrsim t_\mathrm{inflow}$
always holds at all CGM and IGM radii. However, in CR+, $t_\mathrm{cool} \ll
t_\mathrm{inflow}$ at $r\lesssim 360 \,\mathrm{kpc}$. From $t_\mathrm{ff}$ and
$t_\mathrm{inflow}$ we see that CR pressure support in CR+ can mildly delay gas
inflow, but this effect is order-unity. The dominant effect here is that
$t_\mathrm{cool}$ in CR+ is an order of magnitude or more lower than in MHD+.
For infalling gas (especially outside $R_\mathrm{vir}$), this owes to (1) higher
gas densities supported by CR pressure (Fig. \ref{fig:profile.flow.m12i}), (2)
lower temperatures (more dominant inside $R_\mathrm{vir}$, owing to the lack of
a shock), and (3) differences in the thermal phase structure within a given
shell, discussed below.

\subsection{Cooling efficiency with and without dominant CR pressure}
\label{sec:cool_eff}

To better understand why cooling is more efficient in CR+, we derive some
scaling relations for $t_\mathrm{cool}$ as follows. Since thermal pressure
equilibrium is satisfied in MHD+, at any fixed radius, $P \sim n k_B T \sim
\mathrm{constant}$ locally (within a shell at a distance), therefore $n \propto
T^{-1}$ (see \citealt{ji2020properties}). With the definition of
$t_\mathrm{cool}$ and the following very simple power-law approximation of the
cooling function (sufficient for illustrative purposes here)
\begin{align}
  \Lambda(T) \propto
    \begin{cases}
      T \qquad &\text{for $T\lesssim10^5\,\mathrm{K}$}
        \\
      T^{-1} \qquad &\text{for $T\gtrsim10^5\,\mathrm{K}$}
    \end{cases},
    \label{eq:cool_func}
\end{align}
the cooling time in MHD+ should roughly obey
\begin{align}
  t_\mathrm{cool,MHD} \propto
    \begin{cases}
      T^0 &\text{for $T\lesssim10^5\,\mathrm{K}$}
        \\
      T^3 \qquad &\text{for $T\gtrsim10^5\,\mathrm{K}$}
    \end{cases},
\end{align}
at a given radius $r$ \citep{hollenbach1979molecule}. In contrast, in CR+, since
halo gas is supported by CR pressure rather than thermal pressure, the gas
density profile $\rho \sim \rho_\mathrm{eq} =
f(\dot{E}_\mathrm{cr},\tilde{\kappa}, r,...)$, where $\dot{E}_\mathrm{cr}$ and
$\tilde{\kappa}$ are the CR luminosity and the CR effective diffusion
coefficient \citep{ji2020properties}, thus the gas density is independent of
temperature, i.e., $\rho \propto T^0$. Therefore, we expect:
\begin{align}
  t_\mathrm{cool,CR} \propto
    \begin{cases}
      T^{-1} &\text{for $T\lesssim10^5\,\mathrm{K}$}
        \\
      T^2 \qquad &\text{for $T\gtrsim10^5\,\mathrm{K}$}
    \end{cases},
\end{align}
at a given radius $r$.

\begin{figure*}
  \begin{centering}
    \includegraphics[width={0.9\textwidth}]{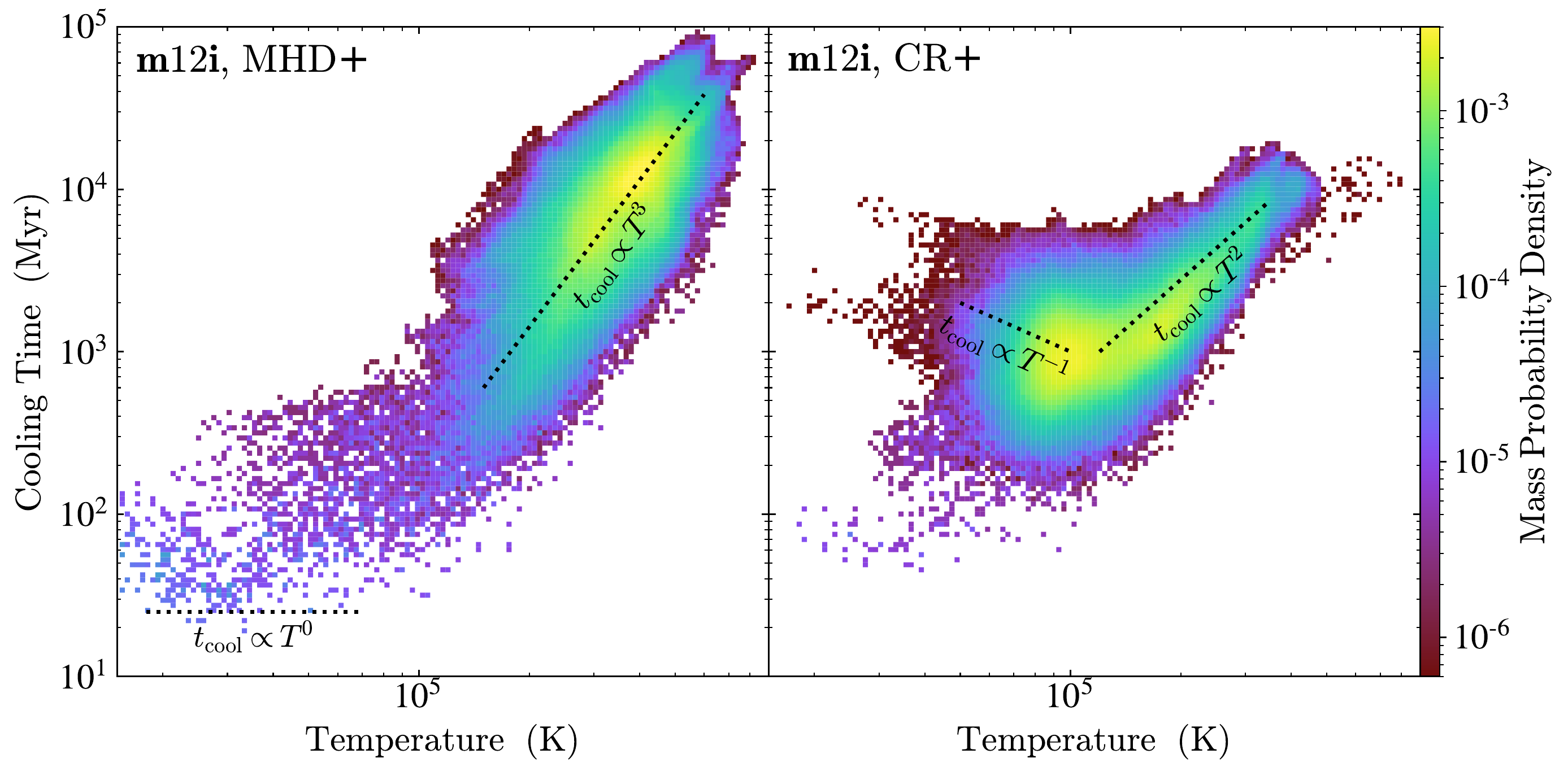}
  \end{centering}
  \caption{Temperature-cooling time phase diagrams of gas in MHD+ (\emph{left})
  and CR+ (\emph{right}) runs of {\bf m12i} at $z = 0$ calculated with the real
  cooling curve. We show gas in a narrow spherical shell at radius $r \approx
  200\,\mathrm{kpc}$, therefore the gas pressure is approximately constant over
  the shell. For gas in warm and hot ($t\gtrsim10^5\,\mathrm{K}$) phases,
  $t_\mathrm{cool}$ in CR+ follows a shallower power law, and thus is greater
  than MHD+ at a given $T$. For gas at a cooler phase ($T\lesssim
  10^5\,\mathrm{K}$), $t_\mathrm{cool}$ in CR+ increases with decreasing
  temperatures, indicating that gas becomes harder to cool after reaching $\sim
  10^5\,\mathrm{K}$ when CR pressure is dominant. The difference of
  $t_\mathrm{cool}$ dependence on $T$ between MHD+ and CR+ owes to the
  difference in temperature $T$ where most of the gas resides, which is
  determined by thermal/CR pressure support -- see more detailed discussions in
  the text. \label{fig:coolingtime.m12i}}
\end{figure*}

We find that these toy relations for $t_\mathrm{cool}$ are consistent with our
simulations, as shown in Fig. \ref{fig:coolingtime.m12i}, where gas within a
shell at $r\sim 200\,\mathrm{kpc}$ is plotted on 2D phase diagrams as a function
of temperature $T$ and cooling time $t_\mathrm{ff}$.\footnote{We also checked
phase diagrams at other radii; although the locations and shapes of the phase
plot change due to radial dependence of density and temperature, the power-law
scalings still hold at different radii.} The difference in this scaling relation
leads to two effects. First, for warm-hot ($T\gtrsim10^5\,\mathrm{K}$) gas,
since $t_\mathrm{cool}$ in MHD+ follows a steeper power ($\propto T^3$) law than
CR+ ($\propto T^2$), gas in MHD+ generally has longer cooling time and thus
stays at a higher temperature range than CR+. Second, for cool
($T\lesssim10^5\,\mathrm{K}$) gas, since $t_\mathrm{cool}\propto T^{-1}$ in CR+,
cooler gas has even longer cooling time, which allows gas to pile up at $T\sim
10^5\,\mathrm{K}$ in CR+.

In greater detail, we investigate how different phases of halo gas contribute to
volume-integrated emissivity $n^2\Lambda$ in our MHD+ and CR+ runs. For MHD+, at
large radii $r\gtrsim 150\,\mathrm{kpc}$, emissivity is dominated by hot,
low-density volume-filling halo gas which cools slowly. While in the inner-CGM
with $50\,\mathrm{kpc} < r< 150\,\mathrm{kpc}$, pressure-confined, high-density
clumps and filaments with an over-density of $\rho > 5 \bar{\rho}(r)$
contributes to $\sim 80\%$ of total emissivity, only filling $~\sim 5\%$ of the
total inner-CGM volume. The rest $\sim 20\%$ emissivity comes from hot and
diffuse halo gas with a volume-filling factor of $\sim 95\%$. For CR+,
emissivity is systematically higher than MHD+ by one order of magnitude, and is
primarily contributed by gas sitting around $\rho \sim \rho_\mathrm{eq}$. This
is also somewhat reflected in Fig. \ref{fig:profile.timescales.m12i}, where both
the magnitude and the dispersion of $t_\mathrm{cool}$ in MHD+ is greater than
those in CR+.

In summary, the gas cooling process evolves differently: in MHD+, cooling gas
condenses out into dense cold clumps, meaning that the volume-filling phase
(relevant for virial shocks) is the reservoir of hot and low-density gas which
cools relatively slowly. In contrast, cooler gas in CR+ can remain at
intermediate temperatures with volume-filling densities. One might worry that
our resolution is insufficient to accurately capture these phenomena
\citep{hummels2019impact}, but recent idealized simulations of the thermal
instability in a patch of the CGM with CRs \citep{butsky2020impact} have
demonstrated exactly the results above, when $P_\mathrm{cr} \gg
P_\mathrm{thermal}$.

\subsection{Heating terms}
\label{sect:heat_cool}

\begin{figure}
  \begin{centering}
    \includegraphics[width={\columnwidth}]{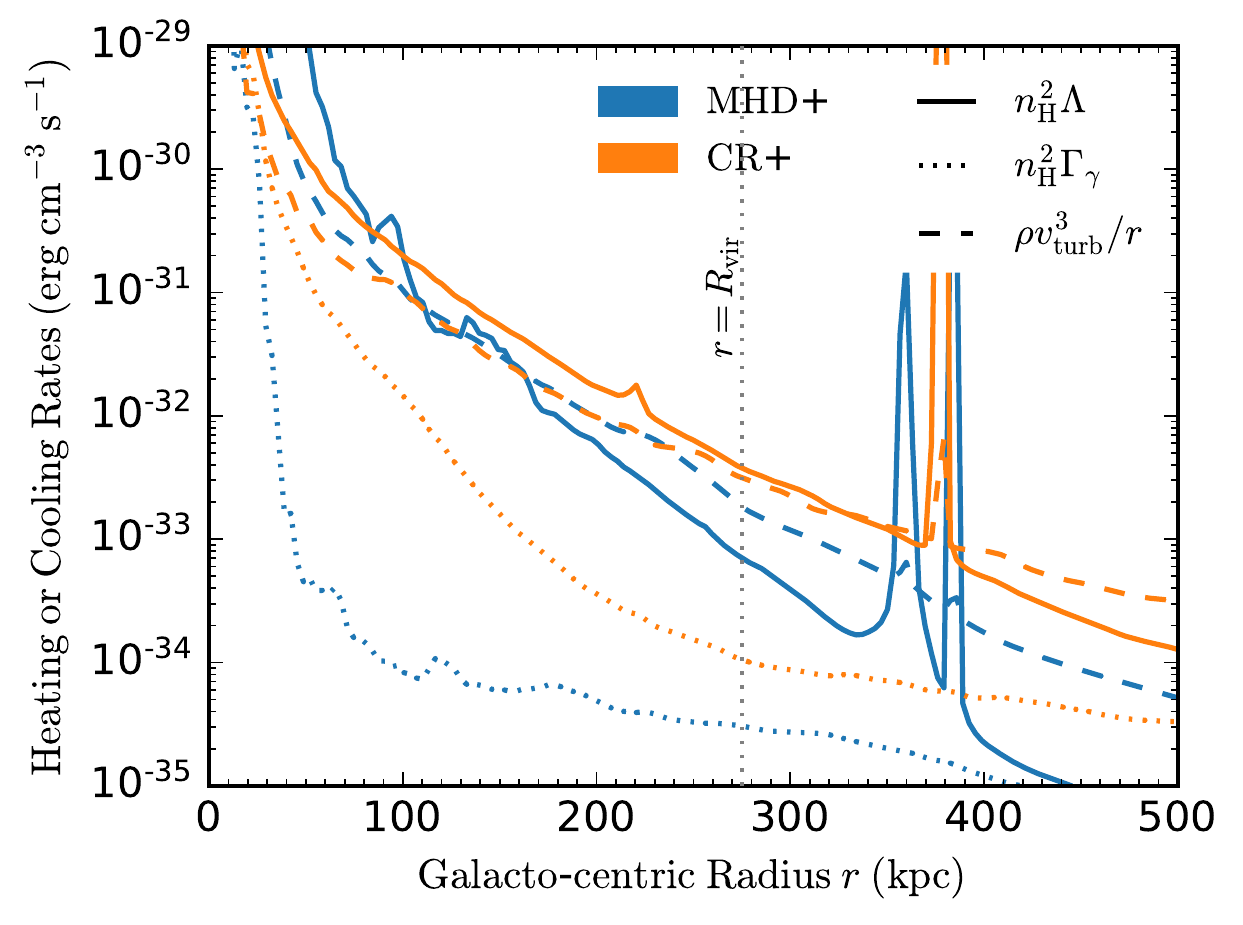}
  \end{centering}
  \caption{Volume-weighted radial profiles of cooling and heating rates in the
  inflow regions: radiative cooling $n^2_\mathrm{H}\Lambda$ (\emph{solid}),
  photon heating $n_\mathrm{H}^2 \Gamma_\gamma$ (\emph{dotted}) and turbulent
  heating $\rho v_\mathrm{turb}^3 / r$ (\emph{dashed}), for the MHD+ (blue) and
  CR+ (orange) runs of the {\bf m12i} halo at $z=0$. Turbulent heating exceeds
  radiative cooling at most radii in MHD+, while is a slightly subdominant
  component inside $R_\mathrm{vir}$ in CR+. Photon heating is
  negligible in both MHD+ and CR+ runs. \label{fig:profile.heat.cool}}
\end{figure}

Fig. \ref{fig:profile.heat.cool} shows volume-averaged profiles of selected
cooling and heating rates for the {\bf m12i} halo runs at $z=0$: radiative
cooling $n^2_\mathrm{H}\Lambda$, photon heating $n_\mathrm{H}^2 \Gamma_\gamma$
and turbulent heating\footnote{Turbulent heating here is estimated as turbulent
energy density $\rho v_\mathrm{turb}^2$ dissipated over an eddy-turnover time
$r/v_\mathrm{turb}$ at a length scale of $r$, with the assumption of homogeneous
and isotropic turbulence which does not strictly apply in our simulations;
therefore the magnitude of turbulent heating quoted here is a rough estimate. In
addition, this estimate actually includes all motion, even large-scale kinetic
motion (e.g., ``eddies'' of $\sim 100\,\mathrm{kpc}$) in the halo as well.}
$\rho v_\mathrm{turb}^3 / r$.  CR non-adiabatic heating (collisional +
streaming) is negligible compared with gas cooling \citep{ji2020properties} and
thus is not shown here. We find that in MHD+, turbulent heating is larger than
radiative cooling at $r \gtrsim 100\,\mathrm{kpc}$. In CR+, turbulent heating is
only slightly subdominant compared to radiative cooling within a radius of
$\lesssim 400\,\mathrm{kpc}$, and exceeds radiative cooling by a factor of a few
at greater radii. Therefore, even though $t_\mathrm{cool}$ is much shorter than
$t_\mathrm{inflow}$ within $R_\mathrm{vir}$ in CR+ (as shown in the right panel
of Fig. \ref{fig:profile.timescales.m12i}), the inflowing gas does not cool to,
but instead stays slightly above the photoionization equilibrium (PIE)
temperature (Fig. 7 in \citealt{ji2020properties}) due to non-negligible
turbulent heating. Photon heating is insignificant compared to gas cooling in
both MHD+ and CR+ runs, as the halo gas is still away from PIE.

\subsection{Halo mass and redshift dependence}
\label{sect:halo_depd}

\subsubsection{Halo Mass}

\begin{figure*}
  \begin{centering}
    \includegraphics[width={\textwidth}]{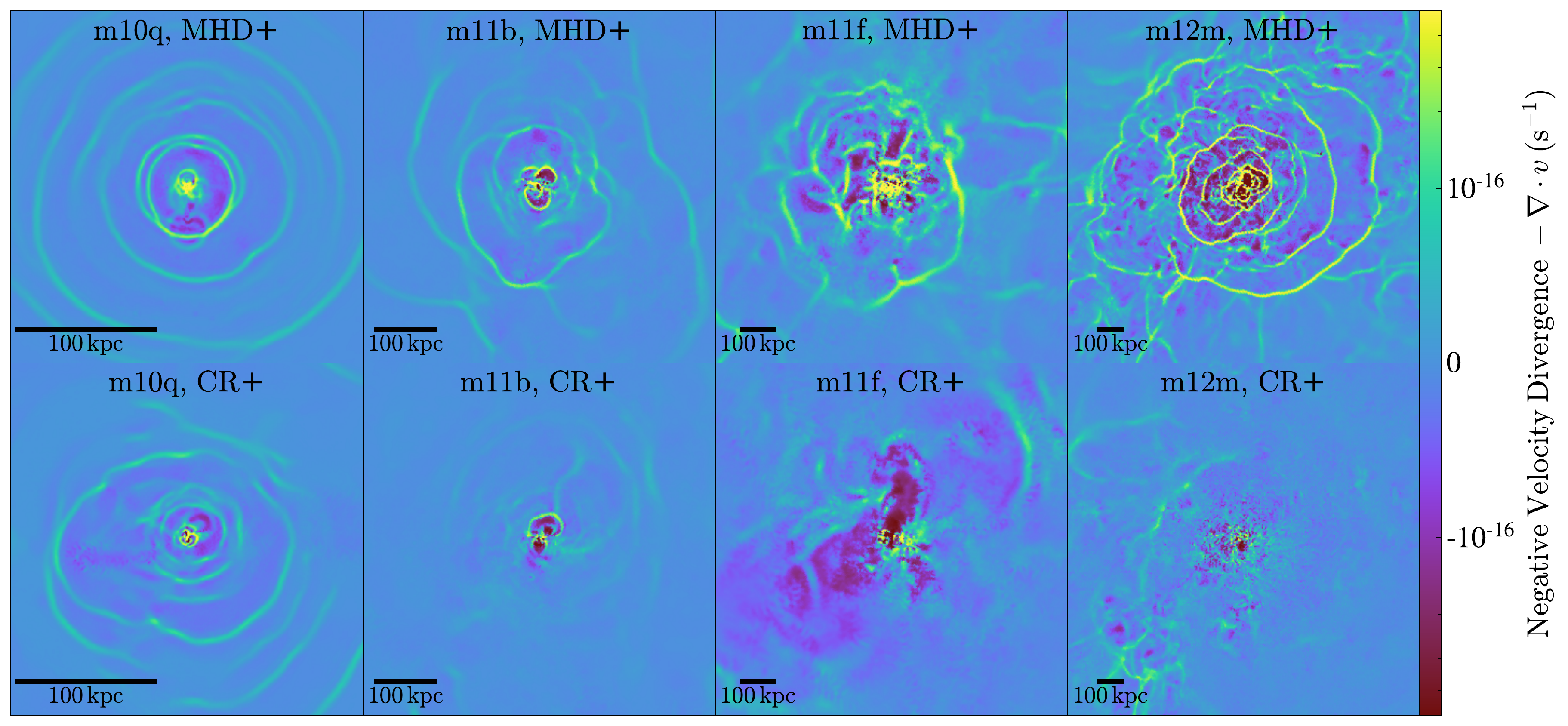}
  \end{centering}
  \caption{Slice plots edge-on to the disk of negative velocity divergences
  $-\nabla\cdot \bm{v}$ (as Fig. \ref{fig:slices_m12i}), in ``MHD+''
  (\emph{top}) and ``CR+'' (\emph{bottom}) runs for {\bf m10q}, {\bf m11b}, {\bf
  m11f} and {\bf m12m} halos at $z=0$. With increasing halo masses, shock
  features becomes more apparent in MHD+, but are more strongly suppressed in
  CR+. Note that shocks at {\bf m10q} are not inflow/virial shocks, but outflow
  shocks from bursts of star-formation feedback. \label{fig:slices.halos}}
\end{figure*}

\begin{figure*}
  \begin{centering}
    \includegraphics[width={0.6\textwidth}]{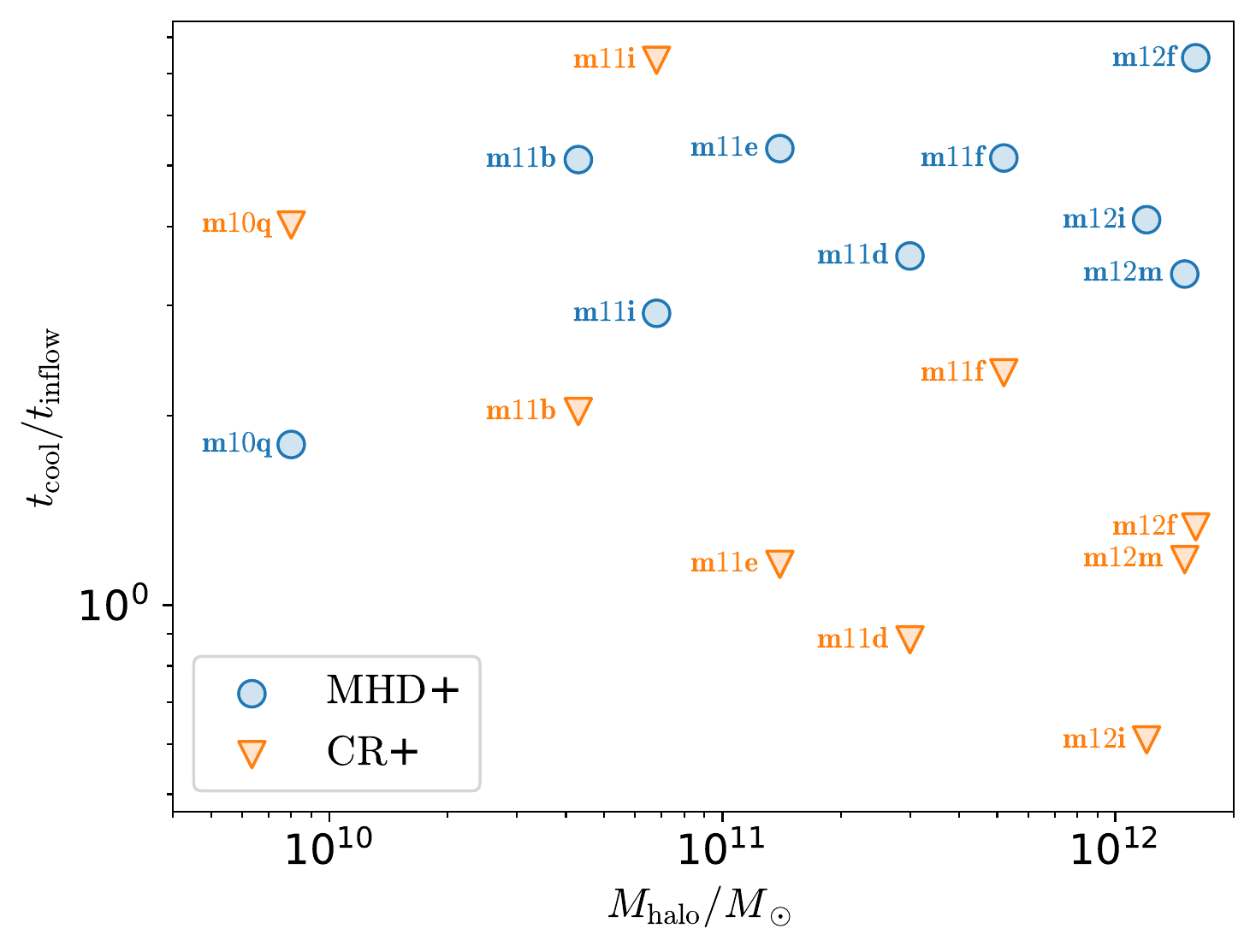}
  \end{centering}
  \caption{$t_\mathrm{cool}/t_\mathrm{inflow}$ versus $M_\mathrm{halo}$ measured
  at $r=R_\mathrm{vir}$ in various halos with mass ranging from
  $10^{10}\,\mathrm{M}_\odot$ to $10^{12}\,\mathrm{M}_\odot$, for both MHD+
  (\emph{blue circles}) and CR+ (\emph{orange triangles}) runs at $z=0$. The CR+
  runs have systematically lower $t_\mathrm{cool}/t_\mathrm{inflow}$ ratios than
  MHD+ in massive halos, becoming chaotic in lower-mass halos where feedback
  from starburst activity is dominant and CR pressure effects are weak (see
  \citealt{hopkins2020but,ji2020properties}). \label{fig:timescale.halos}}
\end{figure*}

As discussed in \S~\ref{sec:intro}, in classic virial shock models (ignoring
feedback, turbulence, magnetic fields, CRs, and assuming spherical symmetry),
one expects something like $t_{\rm cool}/t_{\rm ff} \sim (M_{\rm vir} /
10^{11}\,M_{\odot})^\alpha$ with $\alpha$ between $1/3$ and $2/3$ at $r\sim
R_{\rm vir}$ \citep{silk1977fragmentation,rees1977cooling}, so hot halos only
form at high masses. We therefore extend our investigation to halos of different
masses. Fig.~\ref{fig:slices.halos} shows
visualizations of shocks, for four representative halos of different mass, all
at $z=0$. Fig.~\ref{fig:timescale.halos} summarizes the estimates of $t_{\rm
cool}/t_{\rm inflow}$ around $R_{\rm vir}$ for all our simulations. We note that
since the \emph{volume-filling} phase of the CGM is the most relevant to our
diagnosis of viral shocks, profiles of these timescales are
\emph{volume-averaged} values.

In our least massive halos ($M_{\rm vir} \ll 10^{11}\,M_{\odot}$, e.g. {\bf
m10q}), CRs have very weak effects, as the injection rates are low (owing to
small galaxy masses) and the effects of mechanical feedback from supernovae are
very strong in the shallow potential (this is shown in detail in
\citet{hopkins2020but,ji2020properties}. Naively comparing the ``classic''
virial model it might be surprising that these halos exhibit $t_{\rm cool} \sim
10\,t_{\rm ff}$ and clear shocks (with or without CRs), but these arise from
that same strong stellar feedback\footnote{Even without gas outflows, post-shock
temperatures (at $z\sim 0$) in halos of this mass are low, comparable to the
equilibrium temperature between UVB heating and gas cooling and to the IGM gas
temperature (see discussion in \citealt{kerevs2005galaxies}). Therefore, the
stable shock criterion is satisfied in these objects even if strong shock from
infall is not present.} (the shocks are {\em outward}-propagating, owing to
outflows from successive bursts of star formation, rather than virial shocks;
see \citealt{hopkins2020cosmic}). At slightly higher masses ($\sim
10^{11}\,M_{\odot}$, e.g., {\bf m11b}, {\bf m11i}) we see CRs begin to have an
effect on shock prominence, but the effect is weak.

Above $M_{\rm vir}\gtrsim 10^{11.5}\,M_{\odot}$ (e.g. {\bf m11d}, {\bf m11e},
{\bf m11f}), the effects of CRs become prominent at $\sim R_{\rm vir}$. The
qualitative effects are similar to our {\bf m12i} case study, albeit weaker. In
MHD+, we see that while slightly lower-mass $10^{11}\,M_{\odot}$ halos have
rapidly-dropping $t_{\rm cool} \lesssim t_{\rm inflow}$ inside $R_{\rm vir}$,
these halos are indeed beginning to building up ``hot halo,'' with high central
temperatures and $t_{\rm cool} \sim 5\,t_{\rm inflow} \sim 10\,t_{\rm ff}$ for
the hot volume-filling phase at all radii, and (as a consequence) a visible
quasi-spherical virial shock emerges. In CR+, the combination of larger $t_{\rm
inflow}$ and lower $t_{\rm cool}$ keeps $t_{\rm cool} \sim t_{\rm inflow}$ at
$r\lesssim R_{\rm vir}$. Around $L_{\ast}$ ($M_{\rm vir} \gtrsim
10^{12}\,M_{\odot}$, e.g. {\bf m12i}, {\bf m12m}, {\bf m12f}) the relative
effects of CRs in all cases resembles our {\bf m12i} case study.

Notably, it appears that both the prominence of the hot gas in MHD+ and the CR
pressure at $R_{\rm vir}$ in CR+ build up qualitatively similarly as galaxies
become more massive (e.g.\ we see the radii out to which CRs have a strong
effect, even {\em relative to $R_{\rm vir}$}, increases steadily with $M_{\rm
vir}$). The former owes to the usual cooling physics above. The latter owes to
the relative scaling of the efficiency of star formation (hence the amount of CR
injection, $\dot{E}_{\rm cr}$) increasing in higher-mass halos.

We do not simulate higher-mass halos, because we do not include AGN feedback,
and as a result above $M_{\rm vir} \gg 10^{12}\,M_{\odot}$ these simulations
exhibit severe over-cooling and failure to quench (see e.g.
\citealt{feldmann2015argo,feldmann2017colours,angles2017black,parsotan2020realistic};
Wellons et al., in prep). However, we expect virial shocks to ``return'' even in
CR+ runs at higher masses. If the key criterion is that CR pressure dominate
over thermal or accretion ram pressure at $\sim R_{\rm vir}$, then in the simple
steady-state scalings from \citet{hopkins2020but}, this requires $\dot{E}_{\rm
cr} \gtrsim C\,M_{\rm vir}/\tilde{\kappa}$ where $C$ includes various constants.
If SNe were the primary source of CRs, and star formation were properly quenched
as observed at $M_{\rm vir} \gtrsim 10^{12}\,M_{\odot}$, then we would expect
virial shocks to appear at masses $\gtrsim 10^{12.5}\,M_{\odot}$ or so. However,
AGN can of course also inject CRs. But because (at high masses) $M_{\rm BH}
\propto M_{\ast} \propto M_{\rm vir}^{0.2}$ increases weakly with $M_{\rm vir}$,
it is unlikely that even BHs at high Eddington ratios can lead to CRs dominating
the pressure at $R_{\rm vir}$ in very massive halos with $M_{\rm vir} \gg
10^{13}\,M_{\odot}$ (and simulations like those here including AGN and CRs in
\citet{su2019failure}, we confirm this explicitly).

\subsubsection{Redshift}

\begin{figure*}
  \begin{centering}
    \includegraphics[width={\textwidth}]{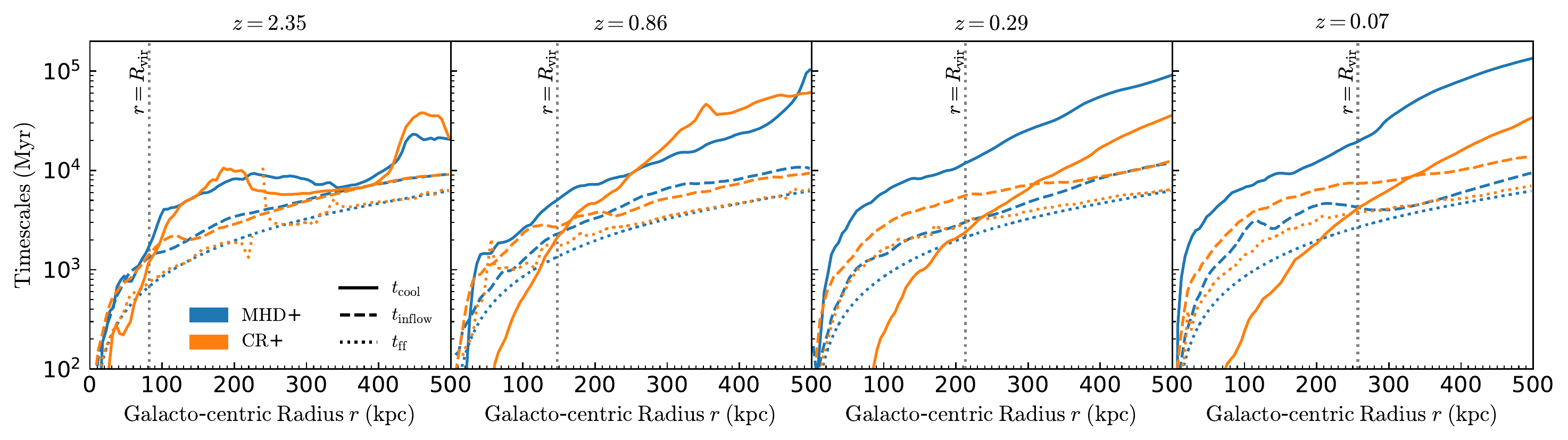}
  \end{centering}
  \caption{Volume-weighted radial profiles of various timescales associated with
  inflows (as Fig. \ref{fig:profile.timescales.m12i}): cooling time
  $t_\mathrm{cool} =3/2\, k T/ n \Lambda$ (\emph{solid}), inflow time
  $t_\mathrm{inflow} = r / v_\mathrm{inflow}$ (\emph{dashed}), freefall time
  $t_\mathrm{ff} = \sqrt{2r/(\nabla \Phi+\nabla P_\mathrm{cr}/\rho)}$
  (\emph{dotted}) in the MHD+ (\emph{blue}) and CR+ (\emph{orange}) runs of {\bf
  m12m} halos at different redshifts of $2.35$, $0.86$, $0.29$ and $0.07$ (where
  $z = 0$ is shown in Fig. \ref{fig:profile.timescales.m12i}), with the
  locations of $R_\mathrm{vir}$ marked with dotted vertical lines.
  \label{fig:profile.timescales.redshift}}
\end{figure*}

\begin{figure*}
    \includegraphics[width={0.981\textwidth},left]{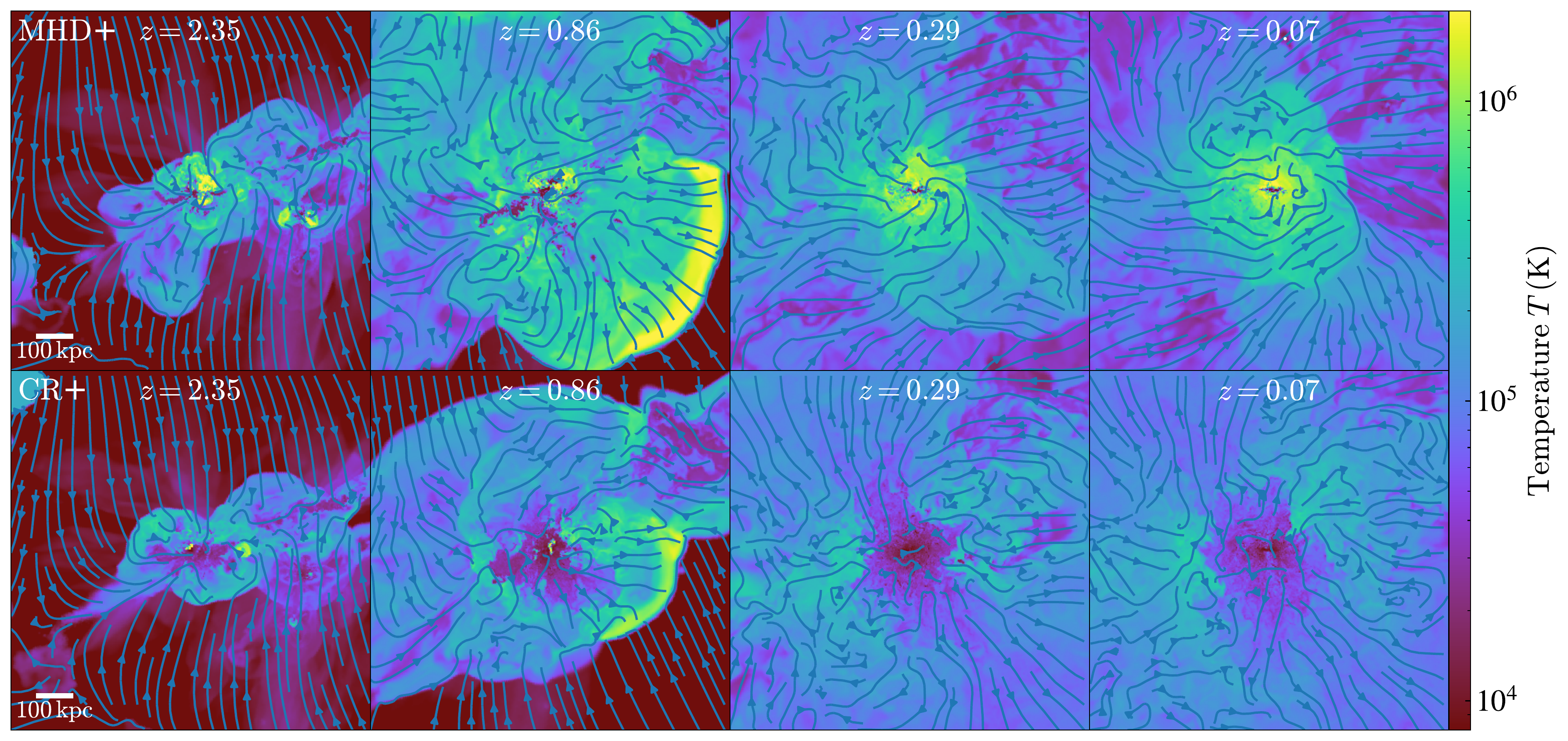}
    \includegraphics[width={\textwidth},left]{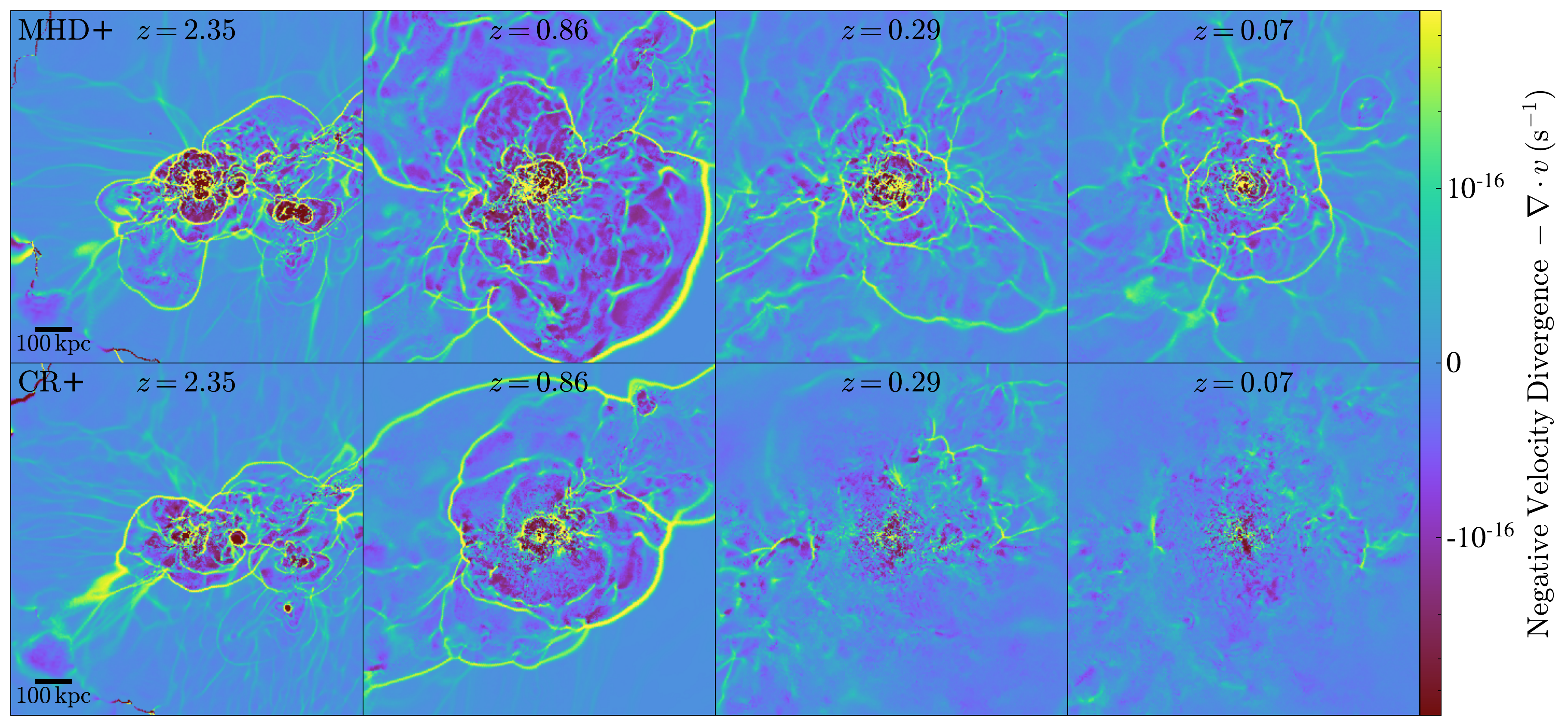}
  \caption{Slice plots edge-on to the disk of temperatures $T$ (with velocity
  streamlines superposed) and negative velocity divergences $-\nabla\cdot
  \bm{v}$ (as Fig. \ref{fig:slices_m12i}), in ``MHD+'' (\emph{top}) and ``CR+''
  (\emph{bottom}) runs for {\bf m12i} halo at different redshifts of $2.35$,
  $0.86$, $0.29$ and $0.07$ ($z = 0$ is shown in Fig. \ref{fig:slices_m12i}). At
  higher redshifts, both MHD+ and CR+ have strong outflowing supernova shocks
  which collide with inflowing gas and heat up the CGM. However, at lower
  redshifts, MHD+ develops a hot halo growing in size with strong virial shocks,
  while in CR+, a cooler halo forms with much less apparent shocks, when CR
  pressure gradually becomes dominant in the CGM. \label{fig:slices.redshift}}
\end{figure*}

We next examine how the effect of CRs on virial shocks evolves over time for a
representative halo {\bf m12i}. As discussed in \citet{hopkins2020but} and
\citet{ji2020properties}, CR pressure is insufficient to balance against gravity
and thus negligible at higher redshifts of $z \gtrsim 1$ -- $2$, therefore it
becomes particularly interesting how CR effects grow with time and affect the
development of virial shocks. Fig. \ref{fig:profile.timescales.redshift} shows
the time evolution of radial profiles of cooling/inflow/freefall times, and Fig.
\ref{fig:slices.redshift} shows visualizations of gas temperatures, velocity
divergences as well as velocity streamlines in {\bf m12i} at different
redshifts.

At a higher redshift $z = 2.35$, no apparent difference in timescales or
temperature/velocity plots between MHD+ and CR+ is observed, since CR effects
are still weak. At $z = 0.86$, CR pressure starts to build up in the inner-CGM
with $r\lesssim 150\,\mathrm{kpc}$ in the CR+ run, which leads to two
consequences: (1) inflow timescale increases since CRs provide pressure support
and delay gas inflowing; (2) cooling timescale becomes shorter than MHD+ since
the volume-filling phase of gas is at lower $T$ and higher $n_\mathrm{H}$, as
discussed in \S\ref{sec:cool_eff}, which corresponds to cooler inner-CGM in CR+
shown in the temperature plot. In the outer-CGM with $r\gtrsim
150\,\mathrm{kpc}$, shocks are prominent in both MHD+ and CR+ at this redshift,
and raise $t_\mathrm{cool}$ by heating up halo gas. Note that these shocks are
actually outflowing supernova shocks rather than virial shocks, which is clearly
demonstrated by outward velocity streamlines near shock fronts in Fig.
\ref{fig:slices.redshift}. SN shocks are slightly less violent in CR+ than MHD+
in terms of shock heating and distances traveled, since at this time SFR in CR+
run is already significantly lower than in MHD+ case so there is less SN energy
available.

At a lower redshift $z = 0.29$, CR effects propagate towards larger radii
$r\gtrsim 150\,\mathrm{kpc}$ and become more substantial. $t_\mathrm{cool}$ in
MHD+ grows significantly from its high-redshift values, in contrast to the CR+
case where the volume-fill phase dominated by CR pressure is at higher
$n_\mathrm{H}$ and lower $T$, and thus $t_\mathrm{cool}$ is lower.
$t_\mathrm{inflow}$ in CR+ is also boosted by CR pressure support, while in MHD+
$t_\mathrm{inflow}$ is systematically smaller due to the lack of pressure
support. At this stage, the ratio $t_\mathrm{cool}/t_\mathrm{inflow}$ becomes
less than $1$ in CR+ for $r\lesssim 350\,\mathrm{kpc}$, and inflowing gas cools
efficiently and feeds the central galaxy gently along the galactic plane,
without creating any prominent shock. CR-driven outflows also start to develop
in the bi-conical regions. In MHD+, $t_\mathrm{cool}$ of the volume-filling
phase is always greater than $t_\mathrm{inflow}$ throughout the halo, and thus
relatively hot gas with long cooling time and insufficient pressure support
continues to fall in and generates virial shocks, building up a hot halo with a
radius of $\sim 100\,\mathrm{kpc}$ at $z = 0.29$. This hot, virialised halo
continues to grow in size, reaching a radius of $\sim 200\,\mathrm{kpc}$ at
$z=0.07$ and $\sim 300\,\mathrm{kpc}$ at $z=0$ (Fig. \ref{fig:slices_m12i});
$t_\mathrm{cool}$ and $t_\mathrm{inflow}$ are further raised respectively by
higher gas temperatures and greater thermal pressure gradients due to virial
shock heating in MHD+.

In short, at higher redshifts $z\gtrsim 2$, there exists very little difference
between MHD+ and CR+, and starburst activity is the governing process that
determines kinematics and thermal states of halo gas. With deceasing redshifts,
CR effects become increasingly more important and develop from the inside out,
reducing $t_\mathrm{cool}$ by modifying phases of inflowing gas and raising
$t_\mathrm{inflow}$ by providing CR pressure support, and thus virial shocks are
almost invariably suppressed in our MW-mass CR+ runs.

\section{Conclusions and discussions}
\label{sec:conclusions}

We explore the effects of CRs on the formation of virial shocks and ``hot
halos'' in galaxies up to $\sim 10^{12}\,M_{\odot}$, in cosmological zoom-in
simulations including star formation, cooling, magnetic fields, conduction,
viscosity, and (optionally) explicit transport and gas coupling between CRs and
gas.

Absent CRs, we show that $\gtrsim 10^{11}\,M_{\odot}$ halos at low redshifts
exhibit clear, quasi-spherical virial shocks at $r \sim R_{\rm vir}$. Within the
virial shock, inflows become well-mixed with thermal-pressure driven confined
outflows, and a classic cooling flow emerges: cooling times of the
volume-filling phase are longer than infall times, infall is subsonic,
temperatures and densities increase towards $r\rightarrow0$. Gas at a given
galacto-centric radius is in approximate thermal pressure equilibrium (so cool
gas is highly overdense).

With CRs as modeled here (under certain assumptions including a streaming speed
at $v_\mathrm{A}$ and a constant diffusion coefficient), the virial shocks are
absent or vastly weaker in $\sim L_{\ast}$ halos when CR pressure dominates over
thermal pressure around $\sim R_{\rm vir}$. Instead of a cooling flow, inflows
develop a simple structure which we show can be analytically predicted assuming
a steady-state, spherical, CR-pressure-dominated halo. Inflows are approximately
isothermal, with constant subsonic inflow velocity, and gas is not in local
thermal pressure equilibrium but rather in {\em total} pressure equilibrium with
CRs constituting most of the pressure. Our analytic model also predicts the
ratio of inflow and outflow densities, velocities and mass fluxes. In this
limit, over-dense gas sinks while under-dense gas is accelerated outwards, while
temperatures are largely set by external heating, so inflows are actually more
dense and have larger thermal pressure compared to outflows. Outflows have
slightly higher velocity while inflows carry slightly larger mass flux, but the
difference is small (tens of percent), where it can be very large absent CRs.

This difference emerges as follows: absent CRs, as a halo grows over time, its
virial temperature and velocity evolve only weakly, but as the virial radius
$R_{\rm vir}$ expands and the mean density of the universe drops, and warm/cool
gas which is thermally unstable precipitates out into over-dense clumps, the
mean gas density $\rho$ of the remaining, inflowing gas at $\gtrsim R_{\rm vir}$
drops\footnote{Note we obtain the same scalings if we consider turnaround or
splashback radii, as these scales similar to $R_{\rm vir}$.} and the cooling time
$t_{\rm cool}(r\sim R_{\rm vir})$ grows until $t_{\rm cool} > t_{\rm inflow}$
at $\sim R_{\rm vir}$ and a stable virial shock forms with a hot halo interior.
CRs modify this in many ways, but we show the most important are (1) the
additional CR pressure can directly support a large weight of gas with low
thermal pressure, and (2) the {\em lack} of local thermal pressure equilibrium
means cooler gas can be volume-filling and is not compressed into dense clumps.
This allows more (factor $\sim$\,a few higher $\sim M_{\rm gas}(<r)$ and
$\rho(r)$) cooler ($\sim 10^{5}\,$K) gas to remain at $r\sim R_{\rm vir}$,
giving rise to order-of-magnitude shorter cooling times $t_{\rm cool}$ at $\sim
R_{\rm vir}$. This maintains $t_{\rm cool} \lesssim t_{\rm inflow}$ within $\sim
R_{\rm vir}$, suppressing the formation of a stable virial shock and hot halo.

From the above, we see CRs do not ``erase'' virial shocks: rather, their effect
is to {\em delay} their formation until galaxies reach later times/larger mass
scales. The key criterion for CRs to have this effect is that CR pressure
dominates over thermal and accretion/ram pressure at $\sim R_{\rm vir}$. This
becomes more challenging in higher-mass halos and at higher redshift, in our
simple steady-state model requiring an injection rate $\dot{E}_{\rm cr} \propto
M_{\rm vir}\,(1+z)^{3} / \tilde{\kappa}$. Knowing that star formation is largely
quenched in galaxies above $M_{\rm vir} \gtrsim 10^{12}\,M_{\odot}$,  even
invoking AGN (with $M_{\rm BH} \propto M_{\ast} \propto M_{\rm vir}^{0.2}$ as
observed at high masses), CR most likely can only ``delay'' the onset of the hot
halo until $M_{\rm vir} \sim 10^{13}\,M_{\odot}$ at $z\sim0$ (with a smaller
effect at higher-$z$), assuming galaxies remain star forming and provide source
of CRs (however most of the observed galaxies are passive in this mass range so
our model does not strictly apply).

Since the thermal state of halo gas in the CR+ runs is quite distinct from MHD+,
we expect dramatically different observables from our CR and non-CR runs. In
addition to ion column densities which are discussed extensively in
\citet{ji2020properties}, we also find different X-ray emissions between MHD+
and CR+. Within a radius of $200\,\mathrm{kpc}$ in {\bf m12i}, the
volume-integrated soft ($0.5$--$2\,\mathrm{keV}$) X-ray luminosity is $\sim
4\times10^{39}\,\mathrm{erg/s}$ in MHD+ and $\sim
6.7\times10^{38}\,\mathrm{erg/s}$ in CR+ respectively; both of these two values
are quite consistent with observations by \citet{li2014chandra} (given SFR and
$M_\star$ in CR+ is a factor of $\sim 3$ smaller than MHD+, Tab. \ref{tbl:sims})
and qualitatively similar to previous FIRE-1 results \citep{van2016impact}. We
note that compared to MHD+, the soft X-ray emission from the CGM (with the
central galaxy excluded) in CR+ is significantly less: the X-ray emission mainly
comes from inflows along the galactic plane, where the CR pressure is less
dominant and thus the gas is warmer compared to the bi-conical outflows. In
contrast, the MHD+ run produces a halo with volume-filling X-ray emissions. We
leave detailed comparisons on X-ray emissions and the SZ effect between MHD+ and
CR+ for future study.

We emphasize, however, our conclusions could be altered by a number of physical
uncertainties, in particular, the uncertainty in CR transport physics. As
discussed in \citet{chan2019cosmic,hopkins2020but}, we deploy a constant CR
diffusion coefficient (with a streaming speed at the Alfv{\'e}n speed
$v_\mathrm{A}$) which is calibrated by CR population modeling in the MW and
$\gamma$-ray observations in nearby galaxies. However, there is no observational
constraint on CR population in the CGM, and it is plausible that the CR
diffusion coefficient is a complicated function of local gas properties, such as
number densities, ionization fractions, magnitudes and morphologies of magnetic
fields, turbulence, etc. (see, e.g., a review paper
\citealt{zweibel2013microphysics}). For instance, when CRs enter the CGM where
gas number densities, magnitudes of magnetic fields and strengths of turbulence
are lower, CRs might become less confined and escape from galaxy halos more
quickly than what we see in our CR+ runs. In \citet{hopkins2020testing} we have
already shown that observationally allowed diffusion coefficients that scales
with physical properties of the gas typically leads to properties of galaxies
intermediate between MHD and CR+. As a consequence, the level of CR pressure
dominance in the CGM and the impact of CRs on the formation of virial shocks in
our simulations might change with more realistic CR transport treatments, which
is subject to future study.

\acknowledgments 
SJ is supported by a Sherman Fairchild Fellowship from Caltech. TKC is supported
by Science and Technology Facilities Council (STFC) astronomy consolidated grant
ST/T000244/1. Support for PFH and co-authors was provided by an Alfred P. Sloan
Research Fellowship, NSF Collaborative Research Grant \#1715847 and CAREER grant
\#1455342, and NASA grants NNX15AT06G, JPL 1589742, 17-ATP17-0214. DK was
supported by NSF grant AST-1715101 and the Cottrell Scholar Award from the
Research Corporation for Science Advancement. CAFG was supported by NSF through
grants AST-1715216 and CAREER award AST-1652522, by NASA through grant
17-ATP17-0067, by STScI through grants HST-GO-14681.011, HST-GO-14268.022-A, and
HST-AR-14293.001-A, and by a Cottrell Scholar Award from the Research
Corporation for Science Advancement. Numerical calculations were run on the
Caltech compute cluster ``Wheeler'', allocations from XSEDE TG-AST120025,
TG-AST130039 and PRAC NSF.1713353 supported by the NSF, and NASA HEC
SMD-16-7592. The data used in this work here were, in part, hosted on facilities
supported by the Scientific Computing Core at the Flatiron Institute, a division
of the Simons Foundation. We have made use of NASA's Astrophysics Data System.
Data analysis and visualization are made with {\small Python 3}, and its
packages including {\small NumPy} \citep{van2011numpy}, {\small SciPy}
\citep{oliphant2007python}, {\small Matplotlib} \citep{hunter2007matplotlib},
and the {\small yt} astrophysics analysis software suite \citep{turk2010yt}, as
well as the spectral simulation code {\small CLOUDY} \citep{ferland20172017}.\\

\dataavailability{The data supporting the plots within this article are
available on reasonable request to the corresponding author. A public version of
the GIZMO code is available at \gizmourl.

Additional data including simulation snapshots, initial conditions, and derived
data products are available at \FIREurl.}\\

\bibliography{ms_extracted}

\begin{thebibliography}{108}
\expandafter\ifx\csname natexlab\endcsname\relax\def\natexlab#1{#1}\fi

\bibitem[{Anderson \& Bregman(2011)}]{anderson2011detection}
Anderson, M.~E., \& Bregman, J.~N. 2011, The Astrophysical Journal, 737, 22

\bibitem[{{Anderson} {et~al.}(2015){Anderson}, {Churazov}, \&
  {Bregman}}]{anderson15}
{Anderson}, M.~E., {Churazov}, E., \& {Bregman}, J.~N. 2015, \mnras, 452, 3905

\bibitem[{Angl{\'e}s-Alc{\'a}zar {et~al.}(2017)Angl{\'e}s-Alc{\'a}zar,
  Faucher-Gigu{\`e}re, Quataert, Hopkins, Feldmann, Torrey, Wetzel, \&
  Kere{\v{s}}}]{angles2017black}
Angl{\'e}s-Alc{\'a}zar, D., Faucher-Gigu{\`e}re, C.-A., Quataert, E., Hopkins,
  P.~F., Feldmann, R., Torrey, P., Wetzel, A., \& Kere{\v{s}}, D. 2017, Monthly
  Notices of the Royal Astronomical Society: Letters, 472, L109

\bibitem[{Birnboim \& Dekel(2003)}]{birnboim2003virial}
Birnboim, Y., \& Dekel, A. 2003, Monthly Notices of the Royal Astronomical
  Society, 345, 349

\bibitem[{{Braginskii}(1965)}]{braginskii:viscosity}
{Braginskii}, S.~I. 1965, Reviews of Plasma Physics, 1, 205

\bibitem[{Brooks {et~al.}(2009)Brooks, Governato, Quinn, Brook, \&
  Wadsley}]{brooks2009role}
Brooks, A., Governato, F., Quinn, T., Brook, C., \& Wadsley, J. 2009, The
  Astrophysical Journal, 694, 396

\bibitem[{{Bryan} \& {Norman}(1998)}]{bryan.norman:1998.mvir.definition}
{Bryan}, G.~L., \& {Norman}, M.~L. 1998, \apj, 495, 80

\bibitem[{Buck {et~al.}(2020)Buck, Pfrommer, Pakmor, Grand, \&
  Springel}]{buck2020effects}
Buck, T., Pfrommer, C., Pakmor, R., Grand, R.~J., \& Springel, V. 2020, Monthly
  Notices of the Royal Astronomical Society, 497, 1712

\bibitem[{Butsky {et~al.}(2020)Butsky, Fielding, Hayward, Hummels, Quinn, \&
  Werk}]{butsky2020impact}
Butsky, I.~S., Fielding, D.~B., Hayward, C.~C., Hummels, C.~B., Quinn, T.~R.,
  \& Werk, J.~K. 2020, arXiv preprint arXiv:2008.04915

\bibitem[{{Cai} {et~al.}(2017){Cai}, {Fan}, {Bian}, {Zabludoff}, {Yang},
  {Prochaska}, {McGreer}, {Zheng}, {Kashikawa}, {Wang}, {Frye}, {Green}, \&
  {Jiang}}]{cai17}
{Cai}, Z., {et~al.} 2017, \apj, 839, 131

\bibitem[{{Cantalupo} {et~al.}(2014){Cantalupo}, {Arrigoni-Battaia},
  {Prochaska}, {Hennawi}, \& {Madau}}]{cantalupo14}
{Cantalupo}, S., {Arrigoni-Battaia}, F., {Prochaska}, J.~X., {Hennawi}, J.~F.,
  \& {Madau}, P. 2014, \nat, 506, 63

\bibitem[{Chan {et~al.}(2019)Chan, Kere{\v{s}}, Hopkins, Quataert, Su, Hayward,
  \& Faucher-Gigu{\`e}re}]{chan2019cosmic}
Chan, T., Kere{\v{s}}, D., Hopkins, P., Quataert, E., Su, K., Hayward, C., \&
  Faucher-Gigu{\`e}re, C. 2019, Monthly Notices of the Royal Astronomical
  Society, 488, 3716

\bibitem[{{Colbrook} {et~al.}(2017){Colbrook}, {Ma}, {Hopkins}, \&
  {Squire}}]{colbrook:passive.scalar.scalings}
{Colbrook}, M.~J., {Ma}, X., {Hopkins}, P.~F., \& {Squire}, J. 2017, \mnras,
  467, 2421

\bibitem[{{Cowie} \& {McKee}(1977)}]{cowie:1977.evaporation}
{Cowie}, L.~L., \& {McKee}, C.~F. 1977, \apj, 211, 135

\bibitem[{Dekel \& Birnboim(2006)}]{dekel2006galaxy}
Dekel, A., \& Birnboim, Y. 2006, Monthly notices of the royal astronomical
  society, 368, 2

\bibitem[{Dijkstra \& Loeb(2009)}]{dijkstra2009lyalpha}
Dijkstra, M., \& Loeb, A. 2009, Monthly Notices of the Royal Astronomical
  Society, 400, 1109

\bibitem[{{Escala} {et~al.}(2018){Escala}, {Wetzel}, {Kirby}, {Hopkins}, {Ma},
  {Wheeler}, {Kere{\v s}}, {Faucher-Gigu{\`e}re}, \&
  {Quataert}}]{escala:turbulent.metal.diffusion.fire}
{Escala}, I., {et~al.} 2018, \mnras, 474, 2194

\bibitem[{Esmerian {et~al.}(2020)Esmerian, Kravtsov, Hafen, Faucher-Giguere,
  Quataert, Stern, Keres, \& Wetzel}]{esmerian2020thermal}
Esmerian, C.~J., Kravtsov, A.~V., Hafen, Z., Faucher-Giguere, C.-A., Quataert,
  E., Stern, J., Keres, D., \& Wetzel, A. 2020, arXiv preprint arXiv:2006.13945

\bibitem[{Fall \& Efstathiou(1980)}]{fall1980formation}
Fall, S.~M., \& Efstathiou, G. 1980, Monthly Notices of the Royal Astronomical
  Society, 193, 189

\bibitem[{Fang {et~al.}(2012)Fang, Bullock, \& Boylan-Kolchin}]{fang2012hot}
Fang, T., Bullock, J., \& Boylan-Kolchin, M. 2012, The Astrophysical Journal,
  762, 20

\bibitem[{Farber {et~al.}(2018)Farber, Ruszkowski, Yang, \&
  Zweibel}]{farber2018impact}
Farber, R., Ruszkowski, M., Yang, H.-Y., \& Zweibel, E. 2018, The Astrophysical
  Journal, 856, 112

\bibitem[{Fardal {et~al.}(2001)Fardal, Katz, Gardner, Hernquist, Weinberg, \&
  Dav{\'e}}]{fardal2001cooling}
Fardal, M.~A., Katz, N., Gardner, J.~P., Hernquist, L., Weinberg, D.~H., \&
  Dav{\'e}, R. 2001, The Astrophysical Journal, 562, 605

\bibitem[{Faucher-Giguere {et~al.}(2016)Faucher-Giguere, Feldmann, Quataert,
  Kere{\v{s}}, Hopkins, \& Murray}]{faucher2016stellar}
Faucher-Giguere, C.-A., Feldmann, R., Quataert, E., Kere{\v{s}}, D., Hopkins,
  P.~F., \& Murray, N. 2016, Monthly Notices of the Royal Astronomical Society:
  Letters, 461, L32

\bibitem[{Faucher-Giguere {et~al.}(2015)Faucher-Giguere, Hopkins, Kere{\v{s}},
  Muratov, Quataert, \& Murray}]{faucher2015neutral}
Faucher-Giguere, C.-A., Hopkins, P.~F., Kere{\v{s}}, D., Muratov, A.~L.,
  Quataert, E., \& Murray, N. 2015, Monthly Notices of the Royal Astronomical
  Society, 449, 987

\bibitem[{Faucher-Gigu{\`e}re \& Kere{\v{s}}(2011)}]{faucher2011small}
Faucher-Gigu{\`e}re, C.-A., \& Kere{\v{s}}, D. 2011, Monthly Notices of the
  Royal Astronomical Society: Letters, 412, L118

\bibitem[{Faucher-Gigu{\`e}re {et~al.}(2010)Faucher-Gigu{\`e}re, Kere{\v{s}},
  Dijkstra, Hernquist, \& Zaldarriaga}]{faucher2010lyalpha}
Faucher-Gigu{\`e}re, C.-A., Kere{\v{s}}, D., Dijkstra, M., Hernquist, L., \&
  Zaldarriaga, M. 2010, The Astrophysical Journal, 725, 633

\bibitem[{Faucher-Gigu{\`e}re {et~al.}(2011)Faucher-Gigu{\`e}re, Kere{\v{s}},
  \& Ma}]{faucher2011baryonic}
Faucher-Gigu{\`e}re, C.-A., Kere{\v{s}}, D., \& Ma, C.-P. 2011, Monthly Notices
  of the Royal Astronomical Society, 417, 2982

\bibitem[{Faucher-Giguere {et~al.}(2009)Faucher-Giguere, Lidz, Zaldarriaga, \&
  Hernquist}]{faucher2009new}
Faucher-Giguere, C.-A., Lidz, A., Zaldarriaga, M., \& Hernquist, L. 2009, The
  Astrophysical Journal, 703, 1416

\bibitem[{Feldmann \& Mayer(2015)}]{feldmann2015argo}
Feldmann, R., \& Mayer, L. 2015, Monthly Notices of the Royal Astronomical
  Society, 446, 1939

\bibitem[{Feldmann {et~al.}(2017)Feldmann, Quataert, Hopkins,
  Faucher-Gigu{\`e}re, \& Kere{\v{s}}}]{feldmann2017colours}
Feldmann, R., Quataert, E., Hopkins, P.~F., Faucher-Gigu{\`e}re, C.-A., \&
  Kere{\v{s}}, D. 2017, Monthly Notices of the Royal Astronomical Society, 470,
  1050

\bibitem[{Ferland {et~al.}(2017)Ferland, Chatzikos, Guzm{\'a}n, Lykins, van
  Hoof, Williams, Abel, Badnell, Keenan, Porter, {et~al.}}]{ferland20172017}
Ferland, G., {et~al.} 2017, arXiv preprint arXiv:1705.10877

\bibitem[{Fielding {et~al.}(2017)Fielding, Quataert, McCourt, \&
  Thompson}]{fielding2017impact}
Fielding, D., Quataert, E., McCourt, M., \& Thompson, T.~A. 2017, Monthly
  Notices of the Royal Astronomical Society, 466, 3810

\bibitem[{Fumagalli {et~al.}(2011)Fumagalli, Prochaska, Kasen, Dekel, Ceverino,
  \& Primack}]{fumagalli2011absorption}
Fumagalli, M., Prochaska, J.~X., Kasen, D., Dekel, A., Ceverino, D., \&
  Primack, J.~R. 2011, Monthly Notices of the Royal Astronomical Society, 418,
  1796

\bibitem[{{Guo} \& {Oh}(2008)}]{guo.oh:cosmic.rays}
{Guo}, F., \& {Oh}, S.~P. 2008, \mnras, 384, 251

\bibitem[{Hafen {et~al.}(2017)Hafen, Faucher-Gigu{\`e}re,
  Angl{\'e}s-Alc{\'a}zar, Kere{\v{s}}, Feldmann, Chan, Quataert, Murray, \&
  Hopkins}]{hafen2017low}
Hafen, Z., {et~al.} 2017, Monthly Notices of the Royal Astronomical Society,
  469, 2292

\bibitem[{{Hahn} \& {Abel}(2011)}]{hahn:2011.music.code.paper}
{Hahn}, O., \& {Abel}, T. 2011, \mnras, 415, 2101

\bibitem[{{Hennawi} {et~al.}(2015){Hennawi}, {Prochaska}, {Cantalupo}, \&
  {Arrigoni-Battaia}}]{hennawi15}
{Hennawi}, J.~F., {Prochaska}, J.~X., {Cantalupo}, S., \& {Arrigoni-Battaia},
  F. 2015, Science, 348, 779

\bibitem[{{Hernquist}(1990)}]{hernquist:profile}
{Hernquist}, L. 1990, \apj, 356, 359

\bibitem[{Hollenbach \& McKee(1979)}]{hollenbach1979molecule}
Hollenbach, D., \& McKee, C.~F. 1979, The Astrophysical Journal Supplement
  Series, 41, 555

\bibitem[{{Holman} {et~al.}(1979){Holman}, {Ionson}, \&
  {Scott}}]{holman:1979.cr.streaming.speed}
{Holman}, G.~D., {Ionson}, J.~A., \& {Scott}, J.~S. 1979, \apj, 228, 576

\bibitem[{{Hopkins}(2018)}]{hopkins:radiation.methods}
{Hopkins}, P.~F., e.~a. 2018, \mnras, in preparation

\bibitem[{{Hopkins}(2015)}]{hopkins:gizmo}
{Hopkins}, P.~F. 2015, \mnras, 450, 53

\bibitem[{{Hopkins}(2016)}]{hopkins:cg.mhd.gizmo}
---. 2016, \mnras, 462, 576

\bibitem[{{Hopkins}(2017)}]{hopkins:gizmo.diffusion}
---. 2017, \mnras, 466, 3387

\bibitem[{Hopkins {et~al.}(2020{\natexlab{a}})Hopkins, Chan, Ji, Hummels,
  Keres, Quataert, \& Faucher-Gigu{\`e}re}]{hopkins2020cosmic}
Hopkins, P.~F., Chan, T., Ji, S., Hummels, C., Keres, D., Quataert, E., \&
  Faucher-Gigu{\`e}re, C.-A. 2020{\natexlab{a}}, arXiv preprint
  arXiv:2002.02462

\bibitem[{Hopkins {et~al.}(2014)Hopkins, Kere{\v{s}}, O{\~n}orbe,
  Faucher-Gigu{\`e}re, Quataert, Murray, \& Bullock}]{hopkins2014galaxies}
Hopkins, P.~F., Kere{\v{s}}, D., O{\~n}orbe, J., Faucher-Gigu{\`e}re, C.-A.,
  Quataert, E., Murray, N., \& Bullock, J.~S. 2014, Monthly Notices of the
  Royal Astronomical Society, 445, 581

\bibitem[{{Hopkins} {et~al.}(2013){Hopkins}, {Narayanan}, \&
  {Murray}}]{hopkins:virial.sf}
{Hopkins}, P.~F., {Narayanan}, D., \& {Murray}, N. 2013, \mnras, 432, 2647

\bibitem[{{Hopkins} \& {Raives}(2016)}]{hopkins:mhd.gizmo}
{Hopkins}, P.~F., \& {Raives}, M.~J. 2016, \mnras, 455, 51

\bibitem[{Hopkins {et~al.}(2020{\natexlab{b}})Hopkins, Squire, Chan, Quataert,
  Ji, Keres, \& Faucher-Gigu{\`e}re}]{hopkins2020testing}
Hopkins, P.~F., Squire, J., Chan, T., Quataert, E., Ji, S., Keres, D., \&
  Faucher-Gigu{\`e}re, C.-A. 2020{\natexlab{b}}, arXiv preprint
  arXiv:2002.06211

\bibitem[{{Hopkins} {et~al.}(2018{\natexlab{a}}){Hopkins}, {Wetzel}, {Kere{\v
  s}}, {Faucher-Gigu{\`e}re}, {Quataert}, {Boylan-Kolchin}, {Murray},
  {Hayward}, {Garrison-Kimmel}, {Hummels}, {Feldmann}, {Torrey}, {Ma},
  {Angl{\'e}s-Alc{\'a}zar}, {Su}, {Orr}, {Schmitz}, {Escala}, {Sanderson},
  {Grudi{\'c}}, {Hafen}, {Kim}, {Fitts}, {Bullock}, {Wheeler}, {Chan},
  {Elbert}, \& {Narayanan}}]{hopkins:fire2.methods}
{Hopkins}, P.~F., {et~al.} 2018{\natexlab{a}}, \mnras, 480, 800

\bibitem[{{Hopkins} {et~al.}(2018{\natexlab{b}}){Hopkins}, {Wetzel}, {Kere{\v
  s}}, {Faucher-Gigu{\`e}re}, {Quataert}, {Boylan-Kolchin}, {Murray},
  {Hayward}, \& {El-Badry}}]{hopkins:sne.methods}
---. 2018{\natexlab{b}}, \mnras, 477, 1578

\bibitem[{Hopkins {et~al.}(2020{\natexlab{c}})Hopkins, Chan, Garrison-Kimmel,
  Ji, Su, Hummels, Kere{\v{s}}, Quataert, \&
  Faucher-Gigu{\`e}re}]{hopkins2020but}
Hopkins, P.~F., {et~al.} 2020{\natexlab{c}}, Monthly Notices of the Royal
  Astronomical Society, 492, 3465

\bibitem[{Hummels {et~al.}(2019)Hummels, Smith, Hopkins, O’Shea, Silvia,
  Werk, Lehner, Wise, Collins, \& Butsky}]{hummels2019impact}
Hummels, C.~B., {et~al.} 2019, The Astrophysical Journal, 882, 156

\bibitem[{Hunter(2007)}]{hunter2007matplotlib}
Hunter, J.~D. 2007, Computing in science \& engineering, 9, 90

\bibitem[{Ji {et~al.}(2018)Ji, Oh, \& McCourt}]{ji2018impact}
Ji, S., Oh, S.~P., \& McCourt, M. 2018, Monthly Notices of the Royal
  Astronomical Society, 476, 852

\bibitem[{Ji {et~al.}(2020)Ji, Chan, Hummels, Hopkins, Stern, Kere{\v{s}},
  Quataert, Faucher-Gigu{\`e}re, \& Murray}]{ji2020properties}
Ji, S., {et~al.} 2020, Monthly Notices of the Royal Astronomical Society, 496,
  4221

\bibitem[{{Jiang} \& {Oh}(2018)}]{jiang.oh:2018.cr.transport.m1.scheme}
{Jiang}, Y.-F., \& {Oh}, S.~P. 2018, \apj, 854, 5

\bibitem[{Kere{\v{s}} \& Hernquist(2009)}]{kerevs2009seeding}
Kere{\v{s}}, D., \& Hernquist, L. 2009, The Astrophysical Journal Letters, 700,
  L1

\bibitem[{Kere{\v{s}} {et~al.}(2009)Kere{\v{s}}, Katz, Fardal, Dav{\'e}, \&
  Weinberg}]{kerevs2009galaxies}
Kere{\v{s}}, D., Katz, N., Fardal, M., Dav{\'e}, R., \& Weinberg, D.~H. 2009,
  Monthly Notices of the Royal Astronomical Society, 395, 160

\bibitem[{Kere{\v{s}} {et~al.}(2005)Kere{\v{s}}, Katz, Weinberg, \&
  Dav{\'e}}]{kerevs2005galaxies}
Kere{\v{s}}, D., Katz, N., Weinberg, D.~H., \& Dav{\'e}, R. 2005, Monthly
  Notices of the Royal Astronomical Society, 363, 2

\bibitem[{{Komarov} {et~al.}(2018){Komarov}, {Schekochihin}, {Churazov}, \&
  {Spitkovsky}}]{komarov:whistler.instability.limiting.transport}
{Komarov}, S., {Schekochihin}, A.~A., {Churazov}, E., \& {Spitkovsky}, A. 2018,
  Journal of Plasma Physics, 84, 905840305

\bibitem[{{Kroupa}(2001)}]{kroupa:2001.imf.var}
{Kroupa}, P. 2001, \mnras, 322, 231

\bibitem[{{Krumholz} \& {Gnedin}(2011)}]{krumholz:2011.molecular.prescription}
{Krumholz}, M.~R., \& {Gnedin}, N.~Y. 2011, \apj, 729, 36

\bibitem[{{Kulsrud} \& {Pearce}(1969)}]{kulsrud.1969:streaming.instability}
{Kulsrud}, R., \& {Pearce}, W.~P. 1969, \apj, 156, 445

\bibitem[{{Kulsrud}(2005)}]{kulsrud:plasma.astro.book}
{Kulsrud}, R.~M. 2005, {Plasma physics for astrophysics} (Princeton, N.J. :
  Princeton University Press)

\bibitem[{{Leitherer} {et~al.}(1999)}]{starburst99}
{Leitherer}, C., {et~al.} 1999, \apjs, 123, 3

\bibitem[{Li {et~al.}(2014)Li, Crain, \& Wang}]{li2014chandra}
Li, J.-T., Crain, R.~A., \& Wang, Q.~D. 2014, Monthly Notices of the Royal
  Astronomical Society, 440, 859

\bibitem[{Li {et~al.}(2008)Li, Li, Wang, Irwin, \& Rossa}]{li2008chandra}
Li, J.-T., Li, Z., Wang, Q.~D., Irwin, J.~A., \& Rossa, J. 2008, Monthly
  Notices of the Royal Astronomical Society, 390, 59

\bibitem[{Martin {et~al.}(2015)Martin, Dijkstra, Henry, Soto, Danforth, \&
  Wong}]{martin2015lyalpha}
Martin, C.~L., Dijkstra, M., Henry, A., Soto, K.~T., Danforth, C.~W., \& Wong,
  J. 2015, The Astrophysical Journal, 803, 6

\bibitem[{{McKenzie} \& {Voelk}(1982)}]{mckenzie.voelk:1982.cr.equations}
{McKenzie}, J.~F., \& {Voelk}, H.~J. 1982, \aap, 116, 191

\bibitem[{{Navarro} {et~al.}(1996){Navarro}, {Frenk}, \& {White}}]{nfw:profile}
{Navarro}, J.~F., {Frenk}, C.~S., \& {White}, S.~D.~M. 1996, \apj, 462, 563

\bibitem[{Nelson {et~al.}(2013)Nelson, Vogelsberger, Genel, Sijacki,
  Kere{\v{s}}, Springel, \& Hernquist}]{nelson2013moving}
Nelson, D., Vogelsberger, M., Genel, S., Sijacki, D., Kere{\v{s}}, D.,
  Springel, V., \& Hernquist, L. 2013, Monthly Notices of the Royal
  Astronomical Society, 429, 3353

\bibitem[{Nelson {et~al.}(2019)Nelson, Pillepich, Springel, Pakmor, Weinberger,
  Genel, Torrey, Vogelsberger, Marinacci, \& Hernquist}]{nelson2019first}
Nelson, D., {et~al.} 2019, Monthly Notices of the Royal Astronomical Society,
  490, 3234

\bibitem[{{O{\~n}orbe} {et~al.}(2014){O{\~n}orbe}, {Garrison-Kimmel}, {Maller},
  {Bullock}, {Rocha}, \& {Hahn}}]{onorbe:2013.zoom.methods}
{O{\~n}orbe}, J., {Garrison-Kimmel}, S., {Maller}, A.~H., {Bullock}, J.~S.,
  {Rocha}, M., \& {Hahn}, O. 2014, \mnras, 437, 1894

\bibitem[{Ocvirk {et~al.}(2008)Ocvirk, Pichon, \& Teyssier}]{ocvirk2008bimodal}
Ocvirk, P., Pichon, C., \& Teyssier, R. 2008, Monthly Notices of the Royal
  Astronomical Society, 390, 1326

\bibitem[{Oliphant(2007)}]{oliphant2007python}
Oliphant, T.~E. 2007, Computing in Science \& Engineering, 9, 10

\bibitem[{Oppenheimer {et~al.}(2010)Oppenheimer, Dav{\'e}, Kere{\v{s}}, Fardal,
  Katz, Kollmeier, \& Weinberg}]{oppenheimer2010feedback}
Oppenheimer, B.~D., Dav{\'e}, R., Kere{\v{s}}, D., Fardal, M., Katz, N.,
  Kollmeier, J.~A., \& Weinberg, D.~H. 2010, Monthly Notices of the Royal
  Astronomical Society, 406, 2325

\bibitem[{Parsotan {et~al.}(2020)Parsotan, Cochrane, Hayward, Angles-Alcazar,
  Feldmann, Faucher-Giguere, Wellons, \& Hopkins}]{parsotan2020realistic}
Parsotan, T., Cochrane, R., Hayward, C., Angles-Alcazar, D., Feldmann, R.,
  Faucher-Giguere, C.-A., Wellons, S., \& Hopkins, P. 2020, arXiv preprint
  arXiv:2009.10161

\bibitem[{{Planck Collaboration} {et~al.}(2013){Planck Collaboration}, {Ade},
  {Aghanim}, {Arnaud}, {Ashdown}, {Atrio-Barandela}, {Aumont}, {Baccigalupi},
  {Balbi}, {Banday}, \& et~al.}]{planck13}
{Planck Collaboration} {et~al.} 2013, \aap, 557, A52

\bibitem[{Prochaska {et~al.}(2017)Prochaska, Werk, Worseck, Tripp, Tumlinson,
  Burchett, Fox, Fumagalli, Lehner, Peeples, {et~al.}}]{prochaska2017cos}
Prochaska, J.~X., {et~al.} 2017, The Astrophysical Journal, 837, 169

\bibitem[{Qu \& Bregman(2019)}]{qu2019warm}
Qu, Z., \& Bregman, J.~N. 2019, The Astrophysical Journal, 880, 89

\bibitem[{Rees \& Ostriker(1977)}]{rees1977cooling}
Rees, M.~J., \& Ostriker, J. 1977, Monthly Notices of the Royal Astronomical
  Society, 179, 541

\bibitem[{Salem {et~al.}(2016)Salem, Bryan, \& Corlies}]{salem2016role}
Salem, M., Bryan, G.~L., \& Corlies, L. 2016, Monthly Notices of the Royal
  Astronomical Society, 456, 582

\bibitem[{Silk(1977)}]{silk1977fragmentation}
Silk, J. 1977, The Astrophysical Journal, 211, 638

\bibitem[{{Skilling}(1971)}]{skilling:1971.cr.diffusion}
{Skilling}, J. 1971, \apj, 170, 265

\bibitem[{Spitzer \& H\"arm(1953)}]{spitzer:conductivity}
Spitzer, L., \& H\"arm, R. 1953, Phys. Rev., 89, 977

\bibitem[{{Squire} {et~al.}(2017{\natexlab{a}}){Squire}, {Kunz}, {Quataert}, \&
  {Schekochihin}}]{squire:2017.kinetic.mhd.alfven}
{Squire}, J., {Kunz}, M.~W., {Quataert}, E., \& {Schekochihin}, A.~A.
  2017{\natexlab{a}}, Physical Review Letters, 119, 155101

\bibitem[{{Squire} {et~al.}(2017{\natexlab{b}}){Squire}, {Quataert}, \&
  {Kunz}}]{squire:2017.max.anisotropy.kinetic.mhd}
{Squire}, J., {Quataert}, E., \& {Kunz}, M.~W. 2017{\natexlab{b}}, Journal of
  Plasma Physics, 83, 905830613

\bibitem[{{Squire} {et~al.}(2017{\natexlab{c}}){Squire}, {Schekochihin}, \&
  {Quataert}}]{squire:2017.max.braginskii.scalings}
{Squire}, J., {Schekochihin}, A.~A., \& {Quataert}, E. 2017{\natexlab{c}}, New
  Journal of Physics, 19, 055005

\bibitem[{Sravan {et~al.}(2016)Sravan, Faucher-Giguere, van~de Voort,
  Kere{\v{s}}, Muratov, Hopkins, Feldmann, Quataert, \&
  Murray}]{sravan2016strongly}
Sravan, N., {et~al.} 2016, Monthly Notices of the Royal Astronomical Society,
  463, 120

\bibitem[{Stern {et~al.}(2018)Stern, Faucher-Gigu{\`e}re, Hennawi, Hafen,
  Johnson, \& Fielding}]{stern2018does}
Stern, J., Faucher-Gigu{\`e}re, C.-A., Hennawi, J.~F., Hafen, Z., Johnson,
  S.~D., \& Fielding, D. 2018, The Astrophysical Journal, 865, 91

\bibitem[{Stern {et~al.}(2016)Stern, Hennawi, Prochaska, \&
  Werk}]{stern2016universal}
Stern, J., Hennawi, J.~F., Prochaska, J.~X., \& Werk, J.~K. 2016, The
  Astrophysical Journal, 830, 87

\bibitem[{Stern {et~al.}(2020)Stern, Faucher-Gigu{\`e}re, Fielding, Quataert,
  Hafen, Gurvich, Ma, Byrne, El-Badry, Angl{\'e}s-Alc{\'a}zar,
  {et~al.}}]{stern2020virialization}
Stern, J., {et~al.} 2020, arXiv preprint arXiv:2006.13976

\bibitem[{Stocke {et~al.}(2013)Stocke, Keeney, Danforth, Shull, Froning, Green,
  Penton, \& Savage}]{stocke2013characterizing}
Stocke, J.~T., Keeney, B.~A., Danforth, C.~W., Shull, J.~M., Froning, C.~S.,
  Green, J.~C., Penton, S.~V., \& Savage, B.~D. 2013, The Astrophysical
  Journal, 763, 148

\bibitem[{{Su} {et~al.}(2017){Su}, {Hopkins}, {Hayward}, {Faucher-Gigu{\`e}re},
  {Kere{\v s}}, {Ma}, \& {Robles}}]{su:2016.weak.mhd.cond.visc.turbdiff.fx}
{Su}, K.-Y., {Hopkins}, P.~F., {Hayward}, C.~C., {Faucher-Gigu{\`e}re}, C.-A.,
  {Kere{\v s}}, D., {Ma}, X., \& {Robles}, V.~H. 2017, \mnras, 471, 144

\bibitem[{Su {et~al.}(2019)Su, Hopkins, Hayward, Ma, Faucher-Gigu{\`e}re,
  Kere{\v{s}}, Orr, Chan, \& Robles}]{su2019failure}
Su, K.-Y., {et~al.} 2019, Monthly Notices of the Royal Astronomical Society,
  487, 4393

\bibitem[{Tumlinson {et~al.}(2017)Tumlinson, Peeples, \&
  Werk}]{tumlinson2017circumgalactic}
Tumlinson, J., Peeples, M.~S., \& Werk, J.~K. 2017, Annual Review of Astronomy
  and Astrophysics, 55, 389

\bibitem[{Turk {et~al.}(2010)Turk, Smith, Oishi, Skory, Skillman, Abel, \&
  Norman}]{turk2010yt}
Turk, M.~J., Smith, B.~D., Oishi, J.~S., Skory, S., Skillman, S.~W., Abel, T.,
  \& Norman, M.~L. 2010, The Astrophysical Journal Supplement Series, 192, 9

\bibitem[{van~de Voort {et~al.}(2020)van~de Voort, Bieri, Pakmor, G{\'o}mez,
  Grand, \& Marinacci}]{van2020effect}
van~de Voort, F., Bieri, R., Pakmor, R., G{\'o}mez, F.~A., Grand, R.~J., \&
  Marinacci, F. 2020, arXiv preprint arXiv:2008.07537

\bibitem[{van~de Voort {et~al.}(2016)van~de Voort, Quataert, Hopkins,
  Faucher-Gigu{\`e}re, Feldmann, Kere{\v{s}}, Chan, \& Hafen}]{van2016impact}
van~de Voort, F., Quataert, E., Hopkins, P.~F., Faucher-Gigu{\`e}re, C.-A.,
  Feldmann, R., Kere{\v{s}}, D., Chan, T., \& Hafen, Z. 2016, Monthly Notices
  of the Royal Astronomical Society, 463, 4533

\bibitem[{van~de Voort \& Schaye(2013)}]{van2013soft}
van~de Voort, F., \& Schaye, J. 2013, Monthly Notices of the Royal Astronomical
  Society, 430, 2688

\bibitem[{van~de Voort {et~al.}(2011)van~de Voort, Schaye, Booth, Haas, \&
  Dalla~Vecchia}]{van2011rates}
van~de Voort, F., Schaye, J., Booth, C., Haas, M.~R., \& Dalla~Vecchia, C.
  2011, Monthly Notices of the Royal Astronomical Society, 414, 2458

\bibitem[{Van Der~Walt {et~al.}(2011)Van Der~Walt, Colbert, \&
  Varoquaux}]{van2011numpy}
Van Der~Walt, S., Colbert, S.~C., \& Varoquaux, G. 2011, Computing in Science
  \& Engineering, 13, 22

\bibitem[{{Wentzel}(1968)}]{wentzel:1968.mhd.wave.cr.coupling}
{Wentzel}, D.~G. 1968, \apj, 152, 987

\bibitem[{Werk {et~al.}(2014)Werk, Prochaska, Tumlinson, Peeples, Tripp, Fox,
  Lehner, Thom, O'Meara, Ford, {et~al.}}]{werk2014cos}
Werk, J.~K., {et~al.} 2014, The Astrophysical Journal, 792, 8

\bibitem[{White \& Rees(1978)}]{white1978core}
White, S.~D., \& Rees, M.~J. 1978, Monthly Notices of the Royal Astronomical
  Society, 183, 341

\bibitem[{{Yan} \&
  {Lazarian}(2008)}]{yan.lazarian.2008:cr.propagation.with.streaming}
{Yan}, H., \& {Lazarian}, A. 2008, \apj, 673, 942

\bibitem[{Zweibel(2013)}]{zweibel2013microphysics}
Zweibel, E.~G. 2013, Physics of Plasmas, 20, 055501

\end{thebibliography}

\end{document}